\documentclass[11pt,a4paper]{article}
\pdfoutput=1
\usepackage{jheppub}

\usepackage{amsmath}
\usepackage{graphicx}
\usepackage{multirow}
\usepackage[utf8]{inputenc}

\newcommand{\ie}{\textit{i.e.,~}}
\newcommand{\eg}{\textit{e.g.,~}}

\newcommand{\Ge}{$^{76}$Ge}
\newcommand{\eps}{\epsilon}
\newcommand{\orcid}[1]{ORCID: \href{http://orcid.org/#1}{#1}}
\usepackage{multicol}

\preprint{
  \begin{flushright}
    FERMILAB-PUB-17-016-T\\
    YITP-SB-17-4\\
    IFT-UAM/CSIC-17-004
  \end{flushright}
}

\title{Curtailing the Dark Side in Non-Standard Neutrino Interactions}

\author[a]{Pilar Coloma}
\affiliation[a]{Theoretical Physics Department, Fermi National Accelerator Laboratory,\\
  P.O.~Box 500, Batavia, IL 60510, U.S.A,}
\emailAdd{pcoloma@fnal.gov}

\author[a,b,1]{Peter B.~Denton,\note{\orcid{0000-0002-5209-872X}}}
\affiliation[b]{Niels Bohr International Academy, University of Copenhagen,\\
  The Niels Bohr Institute, Blegdamsvej 17, DK-2100, Copenhagen, Denmark}
\emailAdd{peterbd1@gmail.com}

\author[c,d,e]{M.~C.~Gonzalez-Garcia,}
\affiliation[c] {Departament de Fis\'{\i}ca Qu\`antica i
  Astrof\'{\i}sica and Institut de Ciencies del Cosmos, Universitat de
  Barcelona, Diagonal 647, E-08028 Barcelona, Spain}
\affiliation[d]{Instituci\'o Catalana de Recerca i Estudis
  Avan\c{c}ats (ICREA), Pg. Lluis Companys 23, 08010 Barcelona,
  Spain.}
\affiliation[e]{C.N.~Yang Institute for Theoretical Physics,
  Stony Brook University, Stony Brook, NY 11794-3840,
  USA}
\emailAdd{maria.gonzalez-garcia@stonybrook.edu}

\author[f]{Michele Maltoni,}
\affiliation[f]{Instituto de F\'{\i}sica Te\'orica UAM/CSIC, Calle de
  Nicol\'as Cabrera 13--15, Universidad Aut\'onoma de Madrid,
  Cantoblanco, E-28049 Madrid, Spain}
\emailAdd{michele.maltoni@csic.es}

\author[g]{Thomas Schwetz}
\affiliation[g]{Institut f\"ur Kernphysik, Karlsruher Institut f\"ur
  Technologie (KIT), D-76021 Karlsruhe, Germany}
\emailAdd{schwetz@kit.edu}

\abstract{In presence of non-standard neutrino interactions the
  neutrino flavor evolution equation is affected by a degeneracy which
  leads to the so-called LMA-Dark solution. It requires a solar mixing
  angle in the second octant and implies an ambiguity in the neutrino
  mass ordering. Non-oscillation experiments are required to break
  this degeneracy.  We perform a combined analysis of data from
  oscillation experiments with the neutrino scattering experiments
  CHARM and NuTeV. We find that the degeneracy can be lifted if the
  non-standard neutrino interactions take place with down quarks, but
  it remains for up quarks. However, CHARM and NuTeV constraints apply
  only if the new interactions take place through mediators not much
  lighter than the electroweak scale. For light mediators we consider
  the possibility to resolve the degeneracy by using data from future
  coherent neutrino-nucleus scattering experiments. We find that, for
  an experiment using a stopped-pion neutrino source, the LMA-Dark
  degeneracy will either be resolved, or the presence of new
  interactions in the neutrino sector will be established with high
  significance.}


\begin{document}

\maketitle

\section{Introduction}

Experiments measuring the flavor composition of solar and atmospheric
neutrinos, as well as neutrinos produced in nuclear reactors and in
accelerators, have established that lepton flavor is not conserved in
neutrino propagation. Instead, it oscillates with a wavelength which
depends on distance and energy, because neutrinos are massive and the
mass states are admixtures of the flavor
states~\cite{Pontecorvo:1967fh, Gribov:1968kq, Maki:1962mu}.  At
present all confirmed oscillation signatures can be well described
with the three flavor neutrinos ($\nu_e$, $\nu_\mu$, $\nu_\tau$) being
quantum superpositions ---~all three with no-vanishing projections~---
of three massive states $\nu_i$ ($i=1,2,3$) with masses $m_i$ leading
to two distinctive splittings (see Ref.~\cite{Esteban:2016qun} for the
latest determination of the neutrino masses and mixings).

Under the assumption that the Standard Model (SM) is the low energy
effective model of a complete high energy theory, neutrino masses
emerge naturally as the first observable consequence from higher
dimensional operators. It is particularly remarkable that the only
dimension five ($d=5$) operator that can be built within the SM
particle content is indeed the Weinberg
operator~\cite{Weinberg:1979sa}, which after electroweak symmetry
breaking leads to a suppression of neutrino masses with the scale of
new physics $\Lambda$, as $m_\nu \sim \mathcal{O}( v^2/\Lambda ) \ll
v$, where $v$ is the Higgs vacuum expectation value.  In this
framework higher dimensional operators may also lead to observable
consequences at low energies in the neutrino sector.  At $d=6$ these
include four-fermion interactions involving neutrinos:
\begin{equation}
  \label{eq:NSI-nc}
  (\bar\nu_\alpha \gamma_\mu P_L \nu_\beta) (\bar f \gamma^\mu P f) \,,
\end{equation}
or
\begin{equation}
  \label{eq:NSI-cc}
  (\bar\nu_\alpha \gamma_\mu P_L \ell_\beta) (\bar f' \gamma^\mu P f) \,,
\end{equation}
where $\alpha,\beta$ are lepton flavor indices, $f,f'$ are SM charged
fermions and $\gamma^\mu$ are the Dirac gamma matrices. Here, $P_L$ is
the left-handed projection operator while $P$ can be either $P_L$ or
$P_R$ (the right-handed projection operator). These operators would
lead to the so-called Non-Standard Interactions (NSI) in the neutrino
sector~\cite{Wolfenstein:1977ue, Valle:1987gv, Guzzo:1991hi} (for
recent reviews, see~\cite{Ohlsson:2012kf, Miranda:2015dra}).  They are
expected to arise generically from the exchange of some mediator state
assumed to be heavier that the characteristic momentum transfer in the
process. Operators in Eq.~\eqref{eq:NSI-cc} lead to the modification
of neutrino production and detection mechanisms via new
charged-current interactions (NSI-CC), while operators in
Eq.~\eqref{eq:NSI-nc} induce new neutral-current processes (NSI-NC).

The operators in Eq.~\eqref{eq:NSI-nc} can modify the forward-coherent
scattering (\ie at zero momentum transfer) of neutrinos as they
propagate through matter via so-called Mikheev-Smirnov-Wolfenstein
(MSW) mechanism~\cite{Wolfenstein:1977ue,
  Mikheev:1986gs}. Consequently their effect can be significantly
enhanced in oscillation experiments where neutrinos travel large
regions of matter, such as is the case for solar and atmospheric
neutrinos.  Indeed, the global analysis of data from oscillation
experiments in the framework of mass induced oscillations in presence
of NSI currently provides some of the strongest constraints on the
size of the NSI affecting neutrino
propagation~\cite{GonzalezGarcia:2011my, Gonzalez-Garcia:2013usa}.

Curiously enough, the analysis from oscillation data still allows for
a window into surprisingly large values of NSI couplings in the the
so-called MSW LMA-Dark (LMA-D)~\cite{Miranda:2004nb} regime. For this
solution, in contrast with the standard MSW LMA regime where the solar
mixing angle is $\theta_{12} \approx 34^\circ$, a value of this mixing
angle in the `dark' octant ($45^\circ<\theta_{12} < 90^\circ$) can fit
solar and reactor data as long as large values of NSI are present,
$\epsilon \sim \mathcal{O}(1)$ where $\eps$ relates the size of the
new physics to the weak interaction.  The origin of this solution is a
degeneracy in oscillation data due to a symmetry of the Hamiltonian
describing neutrino evolution in the presence of
NSI~\cite{GonzalezGarcia:2011my, Gonzalez-Garcia:2013usa,
  Bakhti:2014pva, Coloma:2016gei}. This degeneracy involves not only
the octant of $\theta_{12}$ but also a change in sign of the larger
neutrino mass-squared difference $\Delta m^2_{31}$, which is used to
parameterize the type of neutrino mass ordering, normal versus
inverted. Hence, the LMA-D degeneracy makes it impossible to determine
the neutrino mass ordering by oscillation
experiments~\cite{Coloma:2016gei}, and therefore jeopardizes one of
the main goals of the upcoming neutrino oscillation program.  The only
way to lift the degeneracy is by considering non-oscillation data to
constrain NSI. One goal of this work is to investigate this
possibility.

An alternative way to constrain operators in both
Eqs.~\eqref{eq:NSI-nc} and~\eqref{eq:NSI-cc} is through measurements
of neutrino scattering cross sections with other fermions in the SM.
For a compilation of bounds from scattering experiments on the size of
NSI, see Refs.~\cite{Davidson:2003ha, Biggio:2009nt, Biggio:2009kv}
(notice however that in these studies usually only one NSI parameter
is set to be different from zero at a time).  Generically the
``scattering'' bounds on NSI-CC operators are presently rather
stringent while the bounds on NSI-NC tend to be weaker.  Still, in
Ref.~\cite{Miranda:2004nb} it was found that the combination of
oscillation results with the NSI-NC scattering bounds could
substantially lift the degeneracy between LMA and LMA-D, see
also~\cite{Escrihuela:2009up}.  Nevertheless, a fully combined
analysis of the oscillation data and the relevant results of
scattering experiments is still missing in the literature.  In
particular, it is important to notice that the scattering cross
section measurements are made in the deep-inelastic regime in which,
at difference with the MSW effect in neutrino oscillations, a sizable
momentum is transferred in the interaction.

A novel possibility to study the effect of the NSI operators in
Eq.~\eqref{eq:NSI-nc} is through the measurement of coherent
neutrino-nucleus scattering (CE$\nu$NS).  Several experiments have
been proposed for this task, \eg at a stopped pion
source~\cite{Akimov:2015nza} or at nuclear reactors~\cite{Wong:2008vk,
  Aguilar-Arevalo:2016khx, Aguilar-Arevalo:2016qen, Agnolet:2016zir,
  Billard:2016giu}. In addition, solar neutrinos can in principle
leave a signal in dark matter direct detection
experiments~\cite{Cerdeno:2016sfi}.

In this work, we present the results of a global fit to vector-like
NSI-NC operators, which are those that affect the flavor evolution of
neutrinos in matter, using a combination of oscillation and scattering
data. We will start by presenting the framework of our study in
Sec.~\ref{sec:framework}. We will conclude that not all current
constraints are applicable to all NSI models depending on the mass of
the mediator and the momentum transfer in the interaction. Thus, in
the following sections we will distinguish between two different
classes of models, namely, NSI arising in models with heavy mediators,
and NSI coming from models with light mediators (\ie with masses much
lighter than the electroweak scale).  In both cases the bounds from
oscillation experiments apply and we summarize them in
Sec.~\ref{ssec:osc}. The bounds from scattering experiments in the DIS
regime only apply to models with heavy mediators. For those we
describe our re-analysis of CHARM~\cite{Dorenbosch:1986tb} and
NuTeV~\cite{Zeller:2001hh} data, and the combination with oscillations
in Secs.~\ref{ssec:CHARM}, \ref{ssec:NuTeV},
and~\ref{ssec:prescombheavy}, respectively.  In
Sec.~\ref{sec:COHERENT} we will consider the future sensitivity to NSI
of a coherent neutrino-nucleus scattering experiment, and will take
the COHERENT~\cite{Akimov:2015nza} proposal as an example. In this
case, as in the case of oscillations, the bounds apply to all kind of
models giving rise to NSI at low energies and we present the expected
sensitivity when combined with the present constraints in
Sec.~\ref{sec:cohcomb}, focusing of the possibility to resolve the
LMA-D degeneracy. We summarize our results and conclude in
Sec.~\ref{sec:conclusions}. Analytical considerations related to the
LMA-D degeneracy are presented in appendix~\ref{sec:app}.

\section{The NSI formalism}
\label{sec:framework}

In this work, we will consider NSI affecting neutral-current (NC)
processes relevant to neutrino propagation in matter. The coefficients
accompanying the new operators are usually parametrized in the form:
\begin{equation}
  \label{eq:NSILagrangian}
  \mathcal L_\text{NSI} = -2 \sqrt{2} G_F \sum_{f,P,\alpha,\beta}
  \eps_{\alpha\beta}^{f,P}(\bar\nu_\alpha\gamma^\mu P_L\nu_\beta)
  (\bar f\gamma_\mu P f) \,,
\end{equation}
where $G_F$ is the Fermi constant, $\alpha,\beta$ are flavor indices,
$P\equiv P_L, P_R$ and $f$ is a SM charged fermion. In this notation,
$\epsilon^f_{\alpha\beta}$ parametrizes the strength of the new
interaction with respect to the Fermi constant,
$\epsilon_{\alpha\beta}^f \sim \mathcal{O}(G_X/G_F)$.

\subsection{NSI in neutrino oscillations and the LMA-D degeneracy}

If all possible operators in Eq.~\eqref{eq:NSILagrangian} are added to
the SM Lagrangian, the Hamiltonian of the system which governs
neutrino oscillations in presence of matter is modified as
\begin{equation}
  \label{eq:Hnsi}
  H^\nu = H_\text{vac} +H_\text{mat}
  \equiv \frac{1}{2E} U_\text{vac}
  \begin{pmatrix}
    0 \\ &\Delta m^2_{21} \\ &&\Delta m^2_{31}
  \end{pmatrix}
  U_\text{vac}^\dagger
  + \sqrt{2} G_F N_e(x)
  \begin{pmatrix}
    1+\eps_{ee} & \eps_{e\mu} & \eps_{e\tau} \\
    \eps_{e\mu}^* & \eps_{\mu\mu} & \eps_{\mu\tau} \\
    \eps_{e\tau}^* & \eps_{\mu\tau}^* & \eps_{\tau\tau}
  \end{pmatrix} \,,
\end{equation}
where $U_\text{vac}$ is the 3-lepton mixing matrix in
vacuum~\cite{Pontecorvo:1967fh, Maki:1962mu, Kobayashi:1973fv}.
$N_e(x)$ is the electron density as a function of the distance
traveled by the neutrino in matter. For antineutrinos $H^{\bar\nu} =
(H_\text{vac} - H_\text{mat})^*$.  In Eq.~\eqref{eq:Hnsi} the
generalized matter potential depends on the ``effective'' NSI
parameters $\epsilon_{\alpha\beta}$, defined as
\begin{equation}\label{eq:eps-eff}
  \eps_{\alpha\beta} = \sum_{f=u,d,e} Y_f (x) \eps_{\alpha\beta}^{f,V} \,.
\end{equation}
Note that the sum only extends to those fermions present in the
background medium (up-quarks, down-quarks and electrons), and
$Y_f(x)=N_f(x)/N_e(x)$ is the average ratio of the density for the
fermion $f$ to the density of electrons along the neutrino propagation
path. In the Earth, the ratios $Y_f$ are constant to very good
approximation, while for solar neutrinos they depend on the distance
to the center of the Sun.  The presence of NSI with electrons, $f=e$,
would affect not only neutrino propagation in matter as described in
Eq.~\eqref{eq:Hnsi}, but also the neutrino-electron cross-section in
experiments such as SK, Borexino, and reactor experiments.  Since here
we are only interested in studying the bounds to propagation effects
in what follows we will consider only NSI with quarks. For feasibility
reasons we restrict the analysis to the cases that NSI happen either
for up quarks ($f=u$) or for down quarks ($f=d$).

In principle, the matter potential in Eq.~\eqref{eq:Hnsi} contains a
total of 9 additional parameters per $f$: three diagonal real
parameters, and three off-diagonal complex parameters (\ie 3
additional moduli and 3 complex phases). However, the evolution of the
system given by the Hamiltonian in Eq.~\eqref{eq:Hnsi} is invariant up
to a constant. Therefore, oscillation experiments are only sensitive
to the differences between the diagonal terms in the matter
potential. In what follows, we choose to use the combinations
$\eps_{ee} - \eps_{\mu\mu}$ and $\eps_{\tau\tau} -
\eps_{\mu\mu}$. Also, it should be noted that only vector NSI ($\eps^V
= \eps^L + \eps^R$) contribute to the matter potential in neutrino
oscillations.

As a consequence of the CPT symmetry, neutrino evolution is invariant
if the Hamiltonian in Eq.~\eqref{eq:Hnsi} is transformed as $H^\nu \to
-(H^\nu)^*$, see~\cite{GonzalezGarcia:2011my, Gonzalez-Garcia:2013usa}
for a discussion in the context of NSI.  This transformation can be
realised by changing the oscillation parameters as
\begin{equation}
  \label{eq:osc-deg}
  \begin{aligned}
    \Delta m^2_{31}
    &\to -\Delta m^2_{31} + \Delta m^2_{21} = -\Delta m^2_{32} \,,
    \\
    \sin\theta_{12} &\to \cos\theta_{12} \,,
    \\
    \delta &\to \pi - \delta \,,
  \end{aligned}
\end{equation}
and simultaneously transforming the NSI parameters as
\begin{equation}
  \label{eq:NSI-deg}
  \begin{aligned}
    (\eps_{ee} - \eps_{\mu\mu}) &\to - (\eps_{ee} - \eps_{\mu\mu}) - 2  \,,
    \\
    (\eps_{\tau\tau} - \eps_{\mu\mu}) &\to -(\eps_{\tau\tau} - \eps_{\mu\mu}) \,,
    \\
    \eps_{\alpha\beta} &\to - \eps_{\alpha\beta}^* \quad (\alpha \neq \beta) \,,
  \end{aligned}
\end{equation}
see Refs.~\cite{Gonzalez-Garcia:2013usa, Bakhti:2014pva,
  Coloma:2016gei}.  In Eq.~\eqref{eq:osc-deg}, $\delta$ is the
leptonic Dirac CP phase, and we are using here the parameterization
conventions from Ref.~\cite{Coloma:2016gei}. In Eq.~\eqref{eq:NSI-deg}
we take into account explicitly that oscillation data are only
sensitive to differences in the diagonal elements of the
Hamiltonian. Eq.~\eqref{eq:osc-deg} shows that this degeneracy implies
a change in the octant of $\theta_{12}$ (as manifest in the LMA-D fit
to solar neutrino data~\cite{Miranda:2004nb}) as well as a change in
the neutrino mass ordering, \ie the sign of $\Delta m^2_{31}$. For
that reason it has been called ``generalized mass ordering
degeneracy'' in Ref.~\cite{Coloma:2016gei}.

The $\eps_{\alpha\beta}$ in Eq.~\eqref{eq:NSI-deg} are defined in
Eq.~\eqref{eq:eps-eff} and depend on the density and composition of
the medium. If NSI couple to quarks proportional to charge,
$\eps_{\alpha\beta}^{u,V}=-2\eps_{\alpha\beta}^{d,V}$, they have the
same dependence as the standard matter effect and the degeneracy is
mathematically exact and no combination of oscillation experiments
will be able to resolve it. In this work we consider only NSI with
either up or down quarks and hence the degeneracy will be approximate,
mostly due to the non-trivial neutron density along the neutrino path
inside the Sun~\cite{Gonzalez-Garcia:2013usa}. In particular, the
first transformation in Eq.~\eqref{eq:NSI-deg} becomes
\begin{equation}
  \label{eq:NSI-deg-eff}
  (\eps_{ee}^{q,V} - \eps_{\mu\mu}^{q,V})
  \to - (\eps_{ee}^{q,V} - \eps_{\mu\mu}^{q,V}) - \xi_q \quad (q=u,d) \,,
\end{equation}
where $\xi_q$ depends on the effective matter composition relevant for
the global data and will be determined from the fit.

\subsection{Neutrino scattering and heavy versus light NSI mediators}
\label{sec:scat}

Neutrino scattering experiments are sensitive to different
combinations of $\epsilon_{\alpha\beta}^{f,P}$, depending on whether
the scattering takes place with nuclei or electrons, the number of
protons and neutrons in the target nuclei and other factors. In
Sec.~\ref{sec:current} we will provide the combinations of parameters
constrained by each experiment considered in our global fit.

Before proceeding with the combined analysis let us comment on the
viability of renormalizable models leading to large coefficients in
the neutrino sector. In particular it should be noted that the
operators written in Eq.~\eqref{eq:NSILagrangian} are not gauge
invariant. Once gauge invariance is imposed to the full UV theory, the
NSI operators listed above will be generated together with analogous
operators in the charged lepton sector, which obey the same flavor
structure. In this case, the non-observation of charge lepton flavor
violating processes (CLFV) (\eg $\mu \to eee$) imposes very tight
constraints on the size of neutrino NSI for new physics above the
electroweak (EW) scale.  This eventually renders the effects of NSI
unobservable at neutrino oscillation experiments, unless fine-tuned
cancellations among operators with different dimensions are invoked to
cancel the contributions to CLFV processes. This makes it extremely
challenging to find a model of new physics above the EW scale that can
lead to large NSI effects at low energies, see \eg
Refs.~\cite{Gavela:2008ra, Antusch:2008tz, Wise:2014oea}.

An alternative, studied in some detail in Refs.~\cite{Boehm:2004uq,Farzan:2015doa,
  Farzan:2015hkd, Farzan:2016wym, Forero:2016ghr}, is to assume that
the neutrino NSI are generated by new physics well below the EW
scale. For example, renormalizable, gauge-invariant models leading to
large NSI have been constructed considering a $Z'$ boson associated to
a new $U(1)_X$ symmetry, where $X$ is a certain combination of lepton
or baryon numbers. These models successfully avoid CLFV constraints
through different mechanisms.  Furthermore in these models the
constraints coming from neutrino scattering data such as those from
NuTeV~\cite{Zeller:2001hh} or CHARM
experiments~\cite{Dorenbosch:1986tb} can also be evaded. Generically
the coupling times propagator of the $Z'$ mediating neutrino scattering can be
written as
\begin{equation}
  \frac{g^2}{q^2 - M_{Z'}^2} \,,
\end{equation}
where $q$ is the momentum transfer in the process, and $M_{Z'}$ is the
mass of the new vector boson. It is straightforward to see that, in
the limit $q^2 \gg M_{Z'}^2$, the scattering amplitude will be roughly
proportional to $g^2/q^2 $, leading to a strong suppression of the
effects in deep-inelastic scattering (DIS) experiments for
sufficiently small couplings. Conversely, in neutrino oscillations,
the potential felt by neutrinos in propagation through matter arises
from forward coherent scattering, where the momentum transfer is zero,
leading to effects proportional to $g^2/M_{Z'}^2 $ instead.

Consequently in what follows we will distinguish between the bounds
which apply to two different classes of NSI models:
\begin{enumerate}
\item models with \emph{light} mediators, with masses from
  $\mathcal{O}(10~\text{MeV})$ to $\mathcal{O}(1~\text{GeV})$, and

\item models with \emph{heavy} mediators, with masses from
  $\mathcal{O}(1~\text{GeV})$ to $\mathcal{O}(1~\text{TeV})$.
\end{enumerate}
Those ranges are motivated by the typical energy scales of the
scattering experiments considered below. To illustrate the potential
of a future measurement of coherent neutrino-nucleus scattering we
will consider the COHERENT proposal~\cite{Akimov:2015nza} based on a
stopped pion source, providing neutrinos with energies less than about
50~MeV. The neutrino energies in the CHARM~\cite{Dorenbosch:1986tb}
and NuTeV~\cite{Zeller:2001hh} scattering experiments are $\gtrsim
10$~GeV. Hence, in the light mediator case COHERENT can test NSI,
while effects in CHARM and NuTeV will be suppressed. Conversely, in
the heavy mediator case all bounds would apply.

We do not consider mediators much heavier than
$\mathcal{O}(1~\text{TeV})$ for the following reasons. Since we are
interested mostly in $\epsilon \sim \mathcal{O}(1)$, mediators above
the TeV scale would require large coupling constants violating
perturbativity requirements.  Moreover, in that case the
contact-interaction approximation would hold even at LHC energies, and
the corresponding operators would lead to missing energy
signatures~\cite{Friedland:2011za, Franzosi:2015wha}. A detailed
investigation of this regime is beyond the scope of this work. Note
that for the mediator mass ranges indicated above constraints from LHC
derived under the contact-interaction assumption do not apply.

\section{Current experimental constraints}
\label{sec:current}

Current experimental constraints on vector-like NSI parameters include
those obtained from a global fit to oscillation
data~\cite{Gonzalez-Garcia:2013usa}, as well as those obtained from
results from neutrino scattering data in the deep-inelastic regime. As
mentioned above we will concentrate on NSI with either up or down
quarks.\footnote{ This simplifying assumption should have little impact for the analysis of oscillation data. However, we note that once scattering data are included results may depend on the specific couplings to up and down quarks and, in particular, on whether both couplings are present simultaneously. A general analysis with arbitrary couplings is beyond the scope of this work and will be addressed in the future.}   In this case the most precise scattering results are those
from the CHARM~\cite{Dorenbosch:1986tb} and NuTeV~\cite{Zeller:2001hh}
experiments, which performed $\nu_e$ and $\nu_\mu$ scattering on
nuclei respectively. We present the details of our reanalysis of their
results in Secs.~\ref{ssec:CHARM} and~\ref{ssec:NuTeV} respectively.
As discussed in the previous section we distinguish between NSI from
models with light and heavy mediators. For light mediators, at present
only the bounds from oscillations apply. The bounds for this scenario
are summarized in Sec.~\ref{ssec:osc}.  For models with heavy
mediators both, oscillation and scattering bounds apply and we present
the combined bounds in Sec.~\ref{ssec:prescombheavy}

\subsection{Oscillation experiments}
\label{ssec:osc}

For oscillation constraints on NSI parameters we refer to the
comprehensive global fit in the framework of 3$\nu$ oscillation plus
NSI with up and down quarks performed
in~\cite{Gonzalez-Garcia:2013usa} which we briefly summarize here for
completeness. All oscillation experiments but SNO are only sensitive
to vector NSI-NC via matter effects as described above.  There is some
sensitivity of SNO to axial couplings in their NC data.  For this
reason the analysis in Ref.~\cite{Gonzalez-Garcia:2013usa}, and all
combinations that we will present in what follows are made under the
assumption of purely vector-like NSI.

The fit includes data sets from KamLAND reactor
experiment~\cite{Gando:2010aa} and solar neutrino data from
Chlorine~\cite{Cleveland:1998nv}, Gallex/GNO~\cite{Kaether:2010ag},
SAGE~\cite{Abdurashitov:2009tn},
Super-Kamiokande~\cite{Hosaka:2005um,Cravens:2008aa, Abe:2010hy,
  Smy:2012} Borexino~\cite{Bellini:2011rx,
  Bellini:2008mr} and SNO~\cite{Aharmim:2006kv, Aharmim:2005gt,
  Aharmim:2008kc, Aharmim:2011vm}, together with atmospheric neutrino
results from Super-Kamiokande phases 1--4~\cite{Pik:2012qsy}, LBL results
from MINOS~\cite{Adamson:2013whj, Adamson:2013ue} and
T2K~\cite{Ikeda:2011}, and reactor results from
CHOOZ~\cite{Apollonio:1999ae}, Palo Verde~\cite{Piepke:2002ju}, Double
CHOOZ~\cite{Abe:2012tg}, Daya Bay~\cite{An:2013uza} and
RENO~\cite{Seo:2013}, together with reactor short baseline flux
determination from Bugey~\cite{Declais:1994ma, Declais:1994su},
ROVNO~\cite{Kuvshinnikov:1990ry, Afonin:1988gx},
Krasnoyarsk~\cite{Vidyakin:1987ue, Vidyakin:1994ut},
ILL~\cite{Kwon:1981ua}, G\"osgen~\cite{Zacek:1986cu}, and
SRP~\cite{Greenwood:1996pb}.
  
In principle the analysis depends on the six 3$\nu$ oscillations
parameters plus eight NSI parameters per $f$ target, of which five are
real and three are phases. To keep the fit manageable in
Ref.~\cite{Gonzalez-Garcia:2013usa} only real NSI were considered and
$\Delta m^2_{21}$ effects were neglected in the analysis of
atmospheric and LBL experiments. This renders the analysis 
independent of the CP phase in the leptonic mixing matrix. Furthermore
in Ref.~\cite{Friedland:2004ah} it was shown that strong cancellations
in the oscillation of atmospheric neutrinos occur when two eigenvalues
of $H_\text{mat}$ are equal, and it is for this case that the weakest
constraints are placed. This condition further reduces the parameter
space to the 5 oscillation parameters plus 3 independent NSI
parameters per $f$.

\begin{figure}\centering
  \includegraphics[width=\textwidth]{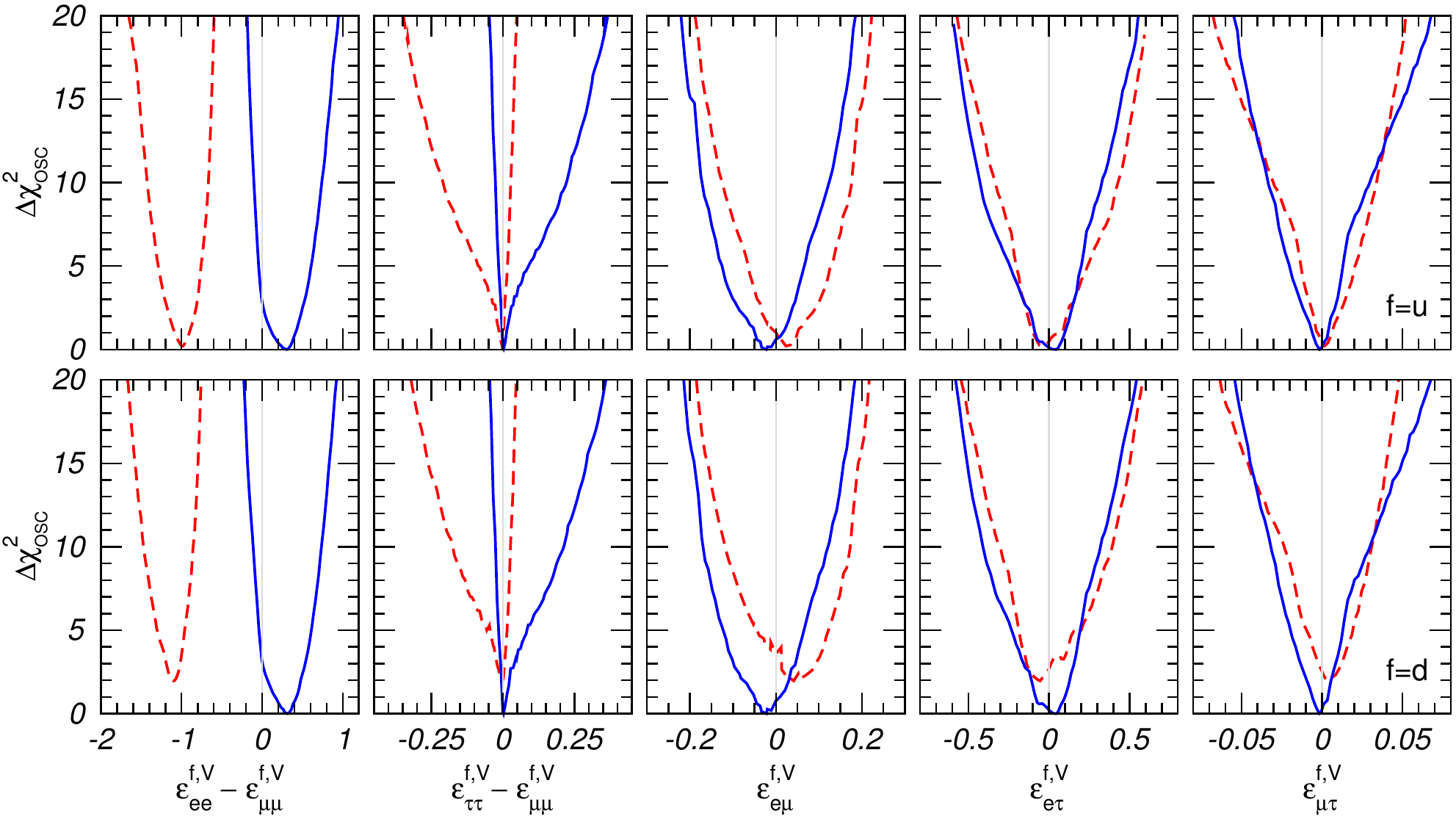}
  \caption{Dependence of the $\Delta\chi^2_\text{OSC}$ function for the
    global analysis of solar, atmospheric, reactor and LBL data on the
    NSI parameters $\eps_{\alpha\beta}^{f,V}$ for $f=u$ (upper panels)
    and $f=d$ (lower panels). Solid curves correspond to the standard
    LMA solution and dashed curves correspond to the LMA-D
    degeneracy. These results correspond to the current limits
    assuming light NSI mediators.  Results adopted from
    Ref.~\cite{Gonzalez-Garcia:2013usa}.}
    \label{fig:nsi-chisq}
\end{figure}

We show in Fig.~\ref{fig:nsi-chisq} the dependence of
$\Delta\chi^2_\text{OSC}$ on each of the relevant NSI coefficients
obtained from the global analysis of oscillation data performed in
Ref.~\cite{Gonzalez-Garcia:2013usa}.  In each panel the results are
shown after marginalization in the full parameter space of oscillation
and considered NSI parameters. In the upper (lower) row these
correspond to vector NSI with up (down) quarks, with all other NSI
(\ie NSI with electrons, axial, and vector ones with down (up) quarks)
set to zero. In each row $\Delta\chi^2_\text{OSC}$ is defined with
respect to the global minimum in the corresponding parameter space.
We also quote the corresponding allowed ranges at 90\%~CL in
Table~\ref{tab:90CL}.

When oscillation parameters are marginalized within the ``standard''
LMA region (solid curves in Fig.~\ref{fig:nsi-chisq}), the global
oscillation analysis slightly favors non-vanishing diagonal NSI, with
the best fit points $\eps^{f,V}_{ee}-\eps^{f,V}_{\mu\mu} = 0.307 \,
(0.316)$ for $f=u(d)$. The reason for this result is the 2$\sigma$
tension in the determination of $\Delta m^2_{21}$ in KamLAND and in
Solar experiments (see, for example~\cite{Esteban:2016qun} for the
latest status on this issue). This tension arises from two facts: i)
neither SNO, SK, nor Borexino shows evidence of the low energy
spectrum turn-up expected in the standard LMA-MSW solution for the
value of $\Delta m^2_{21}$ favored by KamLAND, and ii) the observation
of a non-vanishing day-night asymmetry in SK, whose size is larger
than the one predicted for the $\Delta m^2_{21}$ value indicated of
KamLAND. A small modification of the matter potential reduces this
tension by $\Delta\chi^2\sim 2$. The point of no NSI (all
$\eps_{\alpha\beta}^{f,V} = 0$) has $\Delta
\chi^2_\text{OSC,min}(\text{no NSI}) = 5.4$ (same for up and down
quarks) relative to the best fit, and is allowed at 63\%~CL (for the 5
additional NSI parameters).

The dashed curves in Fig.~\ref{fig:nsi-chisq} are obtained for the
LMA-D degenerate solution~\cite{Miranda:2004nb}, which correspond to
the flipped mass spectrum according to Eq.~\eqref{eq:osc-deg},
including the second octant for $\theta_{12}$. The dashed curves in
the figure clearly follow the transformation from
Eq.~\eqref{eq:NSI-deg-eff}, and comparing the LMA and LMA-D best fit
points for $\eps^{f,V}_{ee}-\eps^{f,V}_{\mu\mu}$ we find
\begin{equation}\label{eq:xi}
  \xi_u = 0.685 \,,\qquad \xi_d = 0.794 \,.
\end{equation}
Although the degeneracy is exact only for $\eps_{\alpha\beta}^{u,V} =
-2\eps_{\alpha\beta}^{d,V}$, we see from the figure that it holds also
to very good accuracy in the case of NSI with up or down quarks
only. We find that for NSI with up-quarks (down-quarks) the LMA-D
solution lies at a $\Delta\chi^2_\text{OSC,min}(\text{LMA-D})\simeq
0.2 (1.9)$.

Since the oscillation results here summarized correspond to the analysis in
Ref.~\cite{Gonzalez-Garcia:2013usa} they do not include data from
oscillation experiments taken since fall 2013.
As discussed above, the LMA-Dark solution emerges from a degeneracy in
the oscillation probability and therefore the inclusion of that additional
oscillation data would not have any quantitative impact relevant to the
conclusions derived below in respect to the status of LMA-Dark. For the same
reason, the LMA-Dark would appear in the analysis of
oscillations if including NSI with general couplings to up
and down quarks. 

As mentioned above, these constraints from oscillations are presently
the only constraints that apply to vector NSI with quarks for models
with a mediator light enough to avoid the bounds from the
deep-inelastic scattering experiments. Conversely for models with
heavier mediators the constraints from DIS experiments apply as we
describe next.

\subsection{CHARM}
\label{ssec:CHARM}

The CHARM collaboration~\cite{Dorenbosch:1986tb} measured the neutral-
and charged-current $\nu_e$ and $\bar\nu_e$ cross sections with
nuclei. To reduce the impact of systematic uncertainties, the ratio of
the neutral-current ($\nu$ plus $\bar\nu$) to charged-current ($\nu$
plus $\bar\nu$) cross sections was reported~\cite{Dorenbosch:1986tb}
\begin{equation}
  R_e = \frac{\sigma(\nu_eN\to\nu_eX) + \sigma(\bar\nu_eN\to\bar\nu_eX)}
  {\sigma(\nu_eN\to eX) + \sigma(\bar\nu_eN\to\bar eX)}
  = 0.406\pm 0.140 \,,
  \label{eq:Re-CHARM}
\end{equation}
which is related to the effective couplings $\tilde g_e^L$ and $\tilde
g_e^R$ for electron neutrinos as
\begin{equation}
  R_e=(\tilde g^L_e)^2+(\tilde g^R_e)^2 \,.
  \label{eq:Re}
\end{equation}
In presence of NSI, the effective couplings read
\begin{equation}
  (\tilde g^P_e)^2=\sum_{q=u,d}
  \left[(g_q^P+\eps_{ee}^{q,P})^2+\sum_{\alpha\neq e}|\eps_{e\alpha}^{q,P}|^2\right]\,,
  \label{eq:gpch}
\end{equation}
where $ g_q^P$ are the SM couplings of the $Z$ boson to quarks, with
tree-level values
\begin{equation}
  \begin{aligned}
    g_u^L &= \frac{1}{2}-\frac{2}{3}\sin^2\theta_W \,,
    &\quad
    g_u^R &= -\frac{2}{3}\sin^2\theta_W \,,
    \\
    g_d^L & =-\frac{1}{2}+\frac{1}{3}\sin^2\theta_W \,,
    &\quad
    g_d^R &= \frac{1}{3}\sin^2\theta_W \,.
  \end{aligned}
  \label{eq:gqP}
\end{equation}
After including one-loop and leading two-loop radiative
corrections~\cite{Erler:2013xha}, they take the values:
$g_u^L=0.3457$, $g_u^R=-0.1553$ $g_d^L=-0.4288$, and $g_d^R=0.0777$
(so $R_{e,\text{SM}}=0.333$), where we have assumed a momentum
transfer $Q^2 \sim 20~\text{GeV}^2$.

\begin{figure}\centering
  \includegraphics[width=3in]{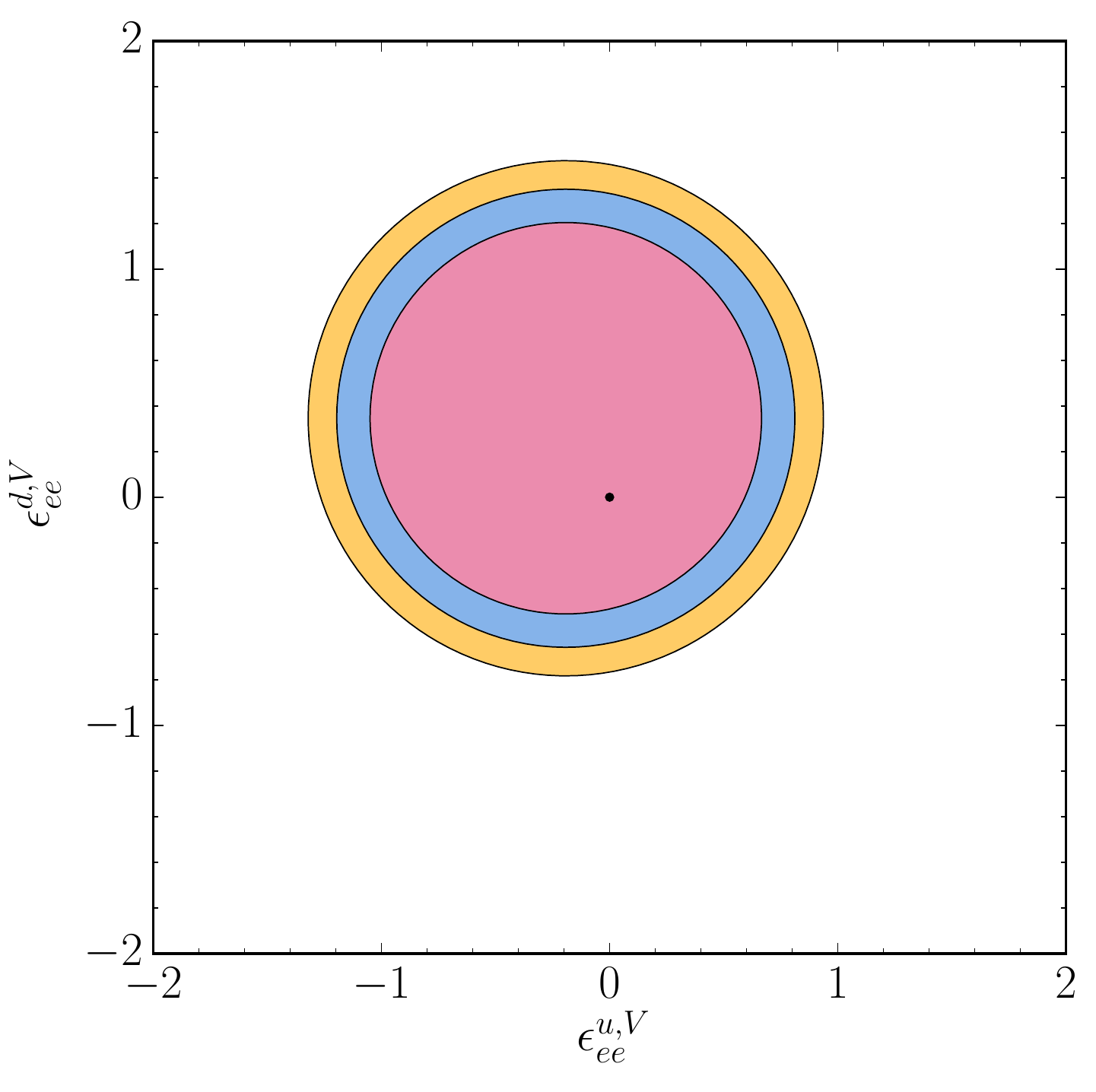}
  \caption{Allowed region from the CHARM measurement on two vector NSI
    parameters, $\eps_{ee}^{u,V}$ and $\eps_{ee}^{d,V}$ (for all other
    NSI couplings set to zero). The black dot shows the SM input
    value, while colored regions correspond to the $1,2,3\sigma$
    contours for two degrees of freedom.}
  \label{fig:CHARM}
\end{figure}

Using the constraint on $R_e$ from Eq.~\eqref{eq:Re-CHARM}, we build
the CHARM contribution to the $\chi^2$ as
\begin{equation}
  \chi^2_\text{CHARM} = \left(\frac{R_{e,\text{NSI}}-R_{e,\text{CHARM}}}{\sigma_\text{CHARM}}\right)^2\,,
\end{equation}
where $R_{e,\text{NSI}}$ is taken from Eq.~\eqref{eq:Re},
$R_{e,\text{CHARM}}=0.406$, and $\sigma_\text{CHARM}=0.140$.  As
illustration we show the bound from CHARM, projected onto the plane
$(\eps_{ee}^{u,V}, \eps_{ee}^{d,V})$ in Fig.~\ref{fig:CHARM} setting
all other NSI couplings to zero. From the figure we see that CHARM
still allows for large $\eps_{ee}^{q,V}$. Also as the SM vector
couplings are $g_q^V=g_q^L+g_q^R\simeq 0.19\, (-0.35)$ for up (down)
quarks, the quadratic and linear contribution of the NSI to $R_e$ have
opposite signs for negative (positive) $\eps^{u,V}_{ee}$
($\eps^{d,V}_{ee}$) and consequently the allowed region extends to
larger negative (positive) values of the corresponding couplings.

\subsection{NuTeV}
\label{ssec:NuTeV}

NuTeV reported measurements of neutral-current (NC) and
charged-current (CC) neutrino-nucleon scattering with both neutrinos
and anti-neutrinos~\cite{Zeller:2001hh}.  The ratios of NC to CC cross
sections for either $\nu$ or $\bar\nu$ scattering from an isoscalar
target can be written as
\begin{equation}
  \begin{aligned}
    R_\mu^\nu
    &= \dfrac{\sigma_\text{NC}(\nu_\mu)}{\sigma_\text{CC}(\nu_\mu)}
    = (\tilde g_\mu^L)^2+r (\tilde g_\mu^R)^2 \,,
    \\
    R_\mu^{\bar\nu}
    &= \dfrac{\sigma_\text{NC}(\bar\nu_\mu)}{\sigma_\text{CC}(\bar\nu_\mu)}
    = (\tilde g_\mu^L)^2+\dfrac {1}{r} (\tilde g_\mu^R)^2 \,,
  \end{aligned}
  \label{eq:rnu}
\end{equation}
with
\begin{equation}
  r = \frac{\sigma_\text{CC}(\bar\nu_\mu)}{\sigma_\text{CC}(\nu_\mu)}\,.
  \label{eq:ratioXsec}
 \end{equation}
In the presence of NSI, the effective couplings $\tilde g_\mu^L$ and
$\tilde g_\mu^R$ get corrected as
\begin{equation}
  g_\text{eff}^{P} \equiv (\tilde g_\mu^P )^2
  = \sum_{q = u,d} \left[ (g_q^P+\eps_{\mu\mu}^{q,P})^2
    +\sum_{\alpha\neq\mu}|\eps_{\mu\alpha}^{q,P}|^2 \right] \,,
  \label{eq:geff}
\end{equation}
where $g_u^P $ and $g_d^P$ are the SM couplings after including
radiative corrections according to the momentum transfer in
NuTeV. Their values can be extracted from the values for the SM
effective couplings $ g_\text{eff,SM}^L$ and $g_\text{eff,SM}^R$ given
in Refs.~\cite{Zeller:2001hh, Zeller:2002he}, which include radiative
corrections.

In order to reconstruct the ratios $R_\mu^{\nu}$ and $R_\mu^{\bar\nu}$
the experiment classifies the events as NC or CC according to the
event length topology. They report their results as ratios of short to
long event rates in either $\nu$ or $\bar\nu$
beams~\cite{Zeller:2001hh}:
\begin{equation}
  \begin{aligned}
    R^\nu_{\mu,\text{exp}}
    &= 0.3916 \pm 0.00069\, (\text{stat}) \pm 0.00044\, (\text{sys})
    \pm 0.0010\, (\text{mod})
    = 0.3919\pm 0.0013 \,,
    \\
    R^{\bar\nu}_{\mu, \text{exp}}
    &= 0.4050 \pm 0.00159\, (\text{stat}) \pm 0.00057\, (\text{sys})
    \pm 0.0021\, (\text{mod})
    = 0.4050\pm 0.0027\,,
    \end{aligned}
  \label{eq:rnutevExp}
\end{equation}
with an overall uncertainty correlation coefficient
$\rho=0.636$~\cite{Zeller:2002he}. The statistical error (stat),
systematic error (sys) and theoretical errors associated to the model
prediction (mod) are indicated separately for convenience.

The reconstructed experimental quantities in Eq.~\eqref{eq:rnutevExp}
cannot be directly compared with the theoretical expression in
Eqs.~\eqref{eq:rnu} and~\eqref{eq:geff} to obtain the constraints on
the NSI, as the relation between the reconstructed short to long event
rates and the cross section ratios in Eq.~\eqref{eq:ratioXsec} can
only be determined using the Monte Carlo of the experiment.  Instead,
one can use the results of the experiment as given in terms of the
fitted effective couplings~\cite{Zeller:2001hh, Zeller:2002he},
\begin{equation}
  (g_\text{eff,exp}^L)^2 = 0.30005\pm 0.00137\,, \quad
  (g_\text{eff,exp}^R)^2 = 0.03076\pm 0.00110\, ,
  \label{eq:geffexp}
\end{equation}
with overall uncertainty correlation coefficient $\rho=-0.017$.  The
NuTeV $\chi^2$ function is then built as
\begin{equation}
  \chi^2_\text{NuTeV} =
  (\vec{X} - \vec{X}_\text{exp})^t V_X^{-1}(\vec{X} -\vec{X}_\text{exp}) \, ,
\end{equation}
where
\begin{equation}
  \vec{X}\equiv
  \begin{pmatrix}
    g^L_\text{eff}\\
    g^R_\text{eff}
  \end{pmatrix} \,,
\end{equation}
and
\begin{equation}
  V_X =
  \begin{pmatrix}
    \sigma(g_\text{eff,exp}^L)^2
    & \sigma(g_\text{eff,exp}^L)\, \sigma(g_\text{eff,exp}^R)\, \rho
    \\
    \sigma(g_\text{eff,exp}^L)\, \sigma(g_\text{eff,exp}^R)\, \rho
    & \sigma(g_\text{eff,exp}^R)^2
  \end{pmatrix} \,,
\end{equation}
is the correlation matrix. Here, $\rho= -0.016$,
$\sigma(g_\text{eff,exp}^L)=0.00137$ and
$\sigma(g_\text{eff,exp}^R)=0.00110$ are taken from
Ref.~\cite{Zeller:2002he}.  Using this $\chi^2$ implementation, one
can easily see that the corresponding SM values for the effective
couplings given in Refs.~\cite{Zeller:2001hh, Zeller:2002he},
$(g_\text{eff,SM}^L)^2=0.3042$ and $(g_\text{eff,SM}^R)^2=0.0301$,
yield a $\chi^2_\text{NuTeV,SM} \sim 9$. This is the well-known NuTeV
anomaly.

Since the publication of the NuTeV results, the requirement of several
additional corrections to their analysis have been pointed
out. Corrections related to nuclear effects, the fact that Fe is not
an isoscalar target and the PDF of the strange quark among
others~\cite{Ball:2009mk, Bentz:2009yy}.  In Ref.~\cite{Bentz:2009yy}
a detailed evaluation of these effects found that all these
corrections shift the central values of the $R$ measurements by
\begin{equation}
  \delta R^\nu_{\mu,\text{exp}} =0.0017\,,\quad
  \delta R^{\bar\nu}_{\mu,\text{exp}}=-0.0016\, ,
\end{equation}
where $\delta R_{\mu,\text{exp}}\equiv R_{\mu,\text{exp\,corr}} -
R_{\mu,\text{exp\,orig}}$.  To translate these shifts into shifts of
the effective couplings we follow the procedure employed by the
collaboration in their fit to the effective couplings by using the
Jacobian $J$ of the transformation between the two sets of variables,
$\delta\vec{X} = J^{-1}\delta\vec{R}$. The Jacobian is determined by
the experimental collaboration via Monte Carlo simulation. Its value
can be found in~\cite{Zeller:2002he}.

With this we get that the results in Eq.~\eqref{eq:geffexp} get
shifted as
\begin{equation}
  \delta g_\text{eff,exp}^L  = 0.00242 \,,
  \quad
  \delta g_\text{eff,exp}^R = -0.00155 \,,
\end{equation}
which bring the corrected experimental results to reasonable agreement with
the SM expectations, $\chi^2_\text{NuTeV,SM}\sim 2.3$.

\begin{figure}\centering
  \includegraphics[width=0.45\textwidth]{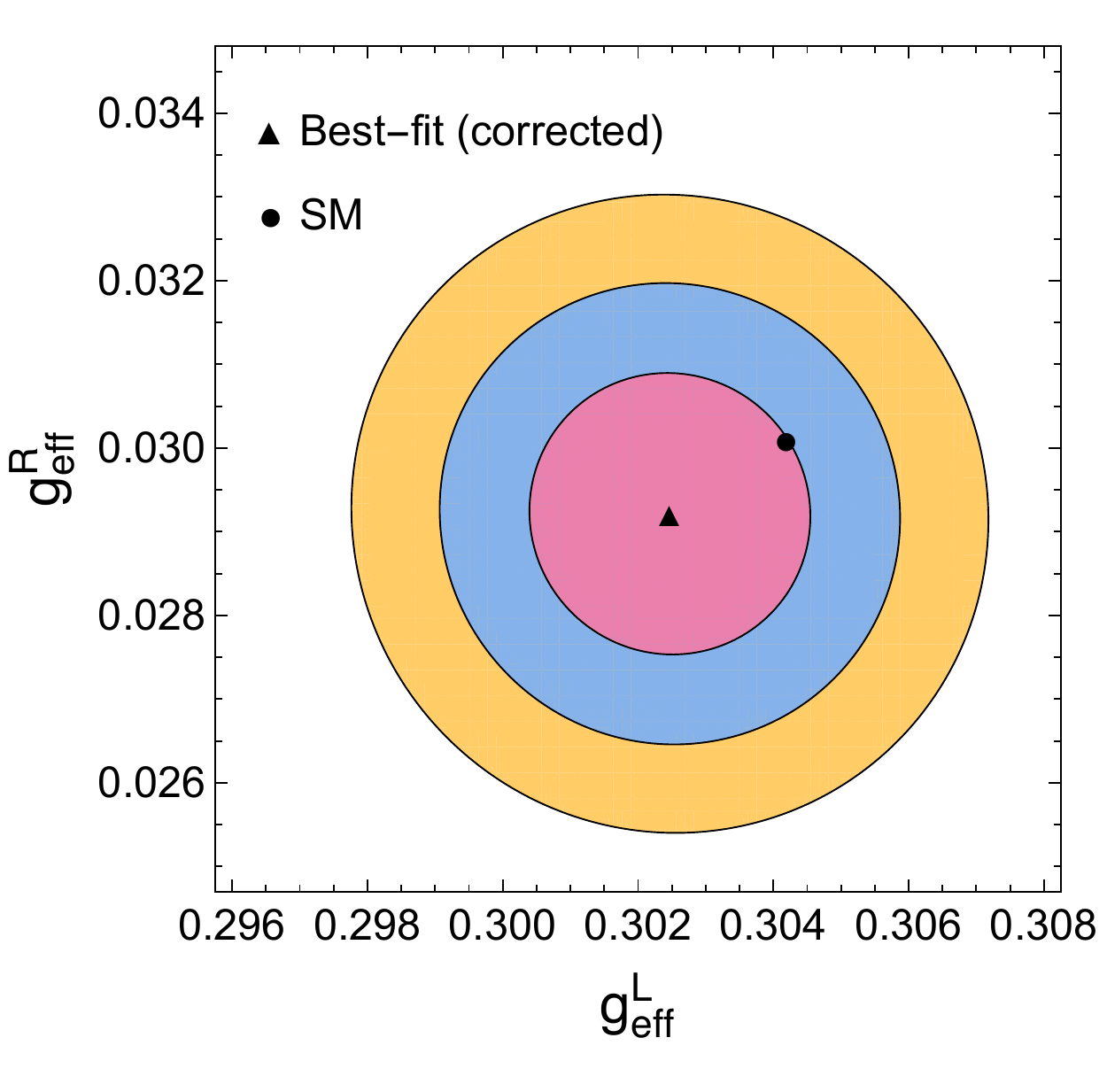}
  \quad
  \includegraphics[width=0.45\textwidth]{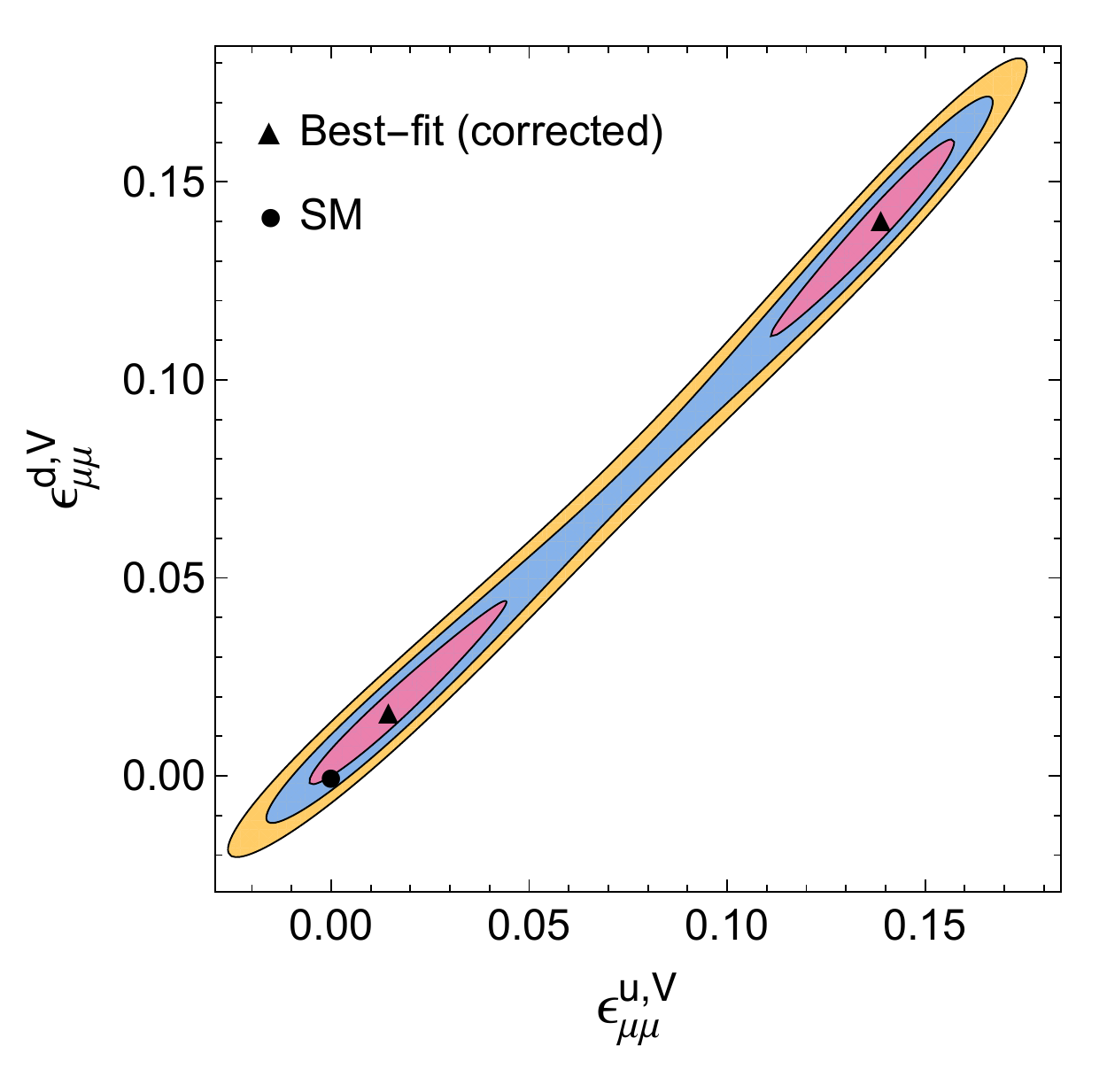}
  \caption{Allowed regions obtained from the NuTeV measurement.
    Contours correspond to $1\sigma$, $2\sigma$ and $3\sigma$ for 2
    degrees of freedom.  The triangle indicates the best-fit points,
    which are completely degenerate. The SM (indicated by a dot) is
    allowed at $\sim 1\sigma$.  The left panel shows allowed regions
    in the plane of the couplings $g^L_\text{eff}$ and
    $g^R_\text{eff}$. The right panel shows the region obtained for
    the two vector NSI parameters $\eps_{\mu\mu}^{u,V}$ and
    $\eps_{\mu\mu}^{d,V}$, after setting all other NSI parameters set
    to zero.}
  \label{fig:NuTeV}
\end{figure}

We adopt these corrected experimental effective coupling results to
derive the corresponding constraints on the NSI, using the
expectations in Eq.~\eqref{eq:geff} with radiative-corrected SM
couplings for the up and down quarks $g_u^L=0.3493$, $g_u^R=-0.1551$
$g_d^L=-0.4269$, and $g_d^R=0.0776$ (which correctly reproduce the SM
predicted values of $(g_\text{eff,SM}^P)^2$ given by the
collaboration).  As illustration, in Fig.~\ref{fig:NuTeV} we show the
bound from NuTeV in the plane of the couplings $g^L_\text{eff}$ and
$g^R_\text{eff}$ (left panel), as well as in the plane
$(\eps_{\mu\mu}^{u,V}, \eps_{\mu\mu}^{d,V})$ (right panel).

\subsection{Global fit to current experiments --- heavy NSI mediators}
\label{ssec:prescombheavy}

With the results above we can now proceed to performed a combined
analysis of the oscillation and scattering experiments by constructing
\begin{equation}
  \chi^2_\text{OSC+SCAT}
  \equiv \chi^2_\text{OSC}+\chi^2_\text{CHARM}+\chi^2_\text{NuTeV}\,,
\end{equation}
and to constrain the vector-like NSI (assuming vanishing axial
couplings) with quarks. These are the present bounds relevant for
models with heavy mediators (as defined in Sec.~\ref{sec:scat}). In
order to keep the analysis feasible, we will consider real NSI
parameters. In all cases, we will show our results as $\Delta \chi^2
\equiv \chi^2 - \chi^2_\text{min}$, where $\chi^2_\text{min}$ is the
global minimum of the $\chi^2$.

\begin{figure}\centering
  \includegraphics[width=\textwidth]{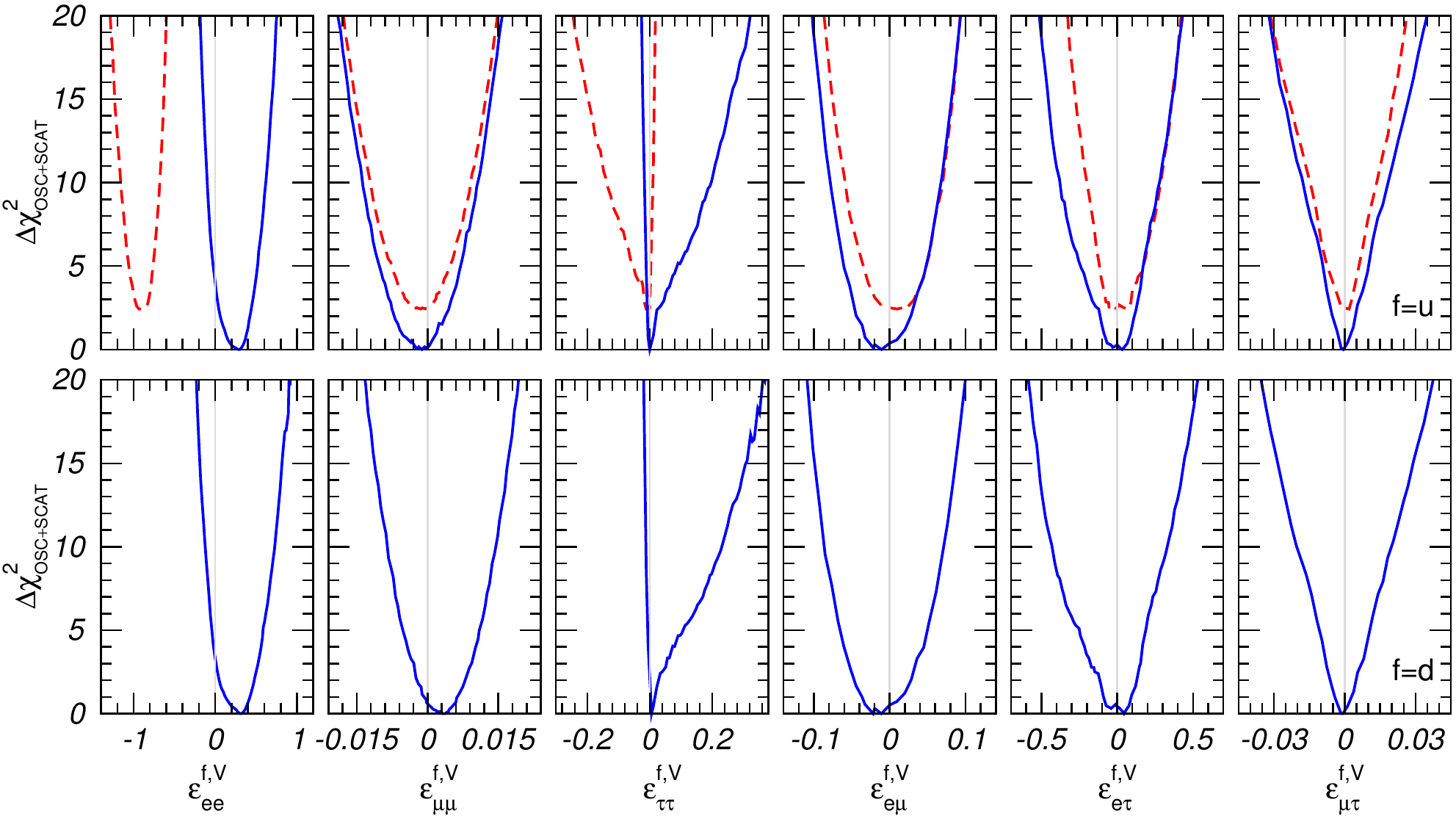}
  \caption{Dependence of the $\Delta\chi^2_\text{OSC+SCAT}$ on the NSI
    parameters $\eps_{\alpha\beta}^{q,V}$ for $q=u$ (upper panels) and
    $q=d$ (lower panels), for both LMA (solid) and LMA-D (dashed)
    regions from the combined analysis of global oscillation and CHARM
    + NuTeV scattering data. These results correspond to the current
    limits assuming heavy NSI mediators.}
  \label{fig:1Dpresent}
\end{figure}

We show in Fig.~\ref{fig:1Dpresent} the dependence of
$\Delta\chi^2_\text{OSC+SCAT}$ on each of the relevant NSI
coefficients (after marginalizing over all oscillation and NSI
undisplayed parameters) for interactions with either up or down
quarks. We also quote the corresponding allowed ranges at 90\%~CL in
Table~\ref{tab:90CL}. From the figure we see how the inclusion of the
results from the scattering experiments resolves the degeneracy for
the flavor-diagonal NSI parameters from oscillations. This results in
very strong bounds in the $\eps_{\mu\mu}^{q,V}$ direction, driven
mostly by NuTeV.\footnote{Note that this strong constraint follows
  from our assumption of interactions either with up or down
  quarks. As visible in Fig.~\ref{fig:NuTeV} there is a strong
  correlation of $\eps_{\mu\mu}^{u,V}$ and $\eps_{\mu\mu}^{d,V}$. In
  the direction $\eps_{\mu\mu}^{u,V} \approx \eps_{\mu\mu}^{d,V}$ the
  NuTeV constraint is much weaker and values up to
  $\eps_{\mu\mu}^{u,V} \approx \eps_{\mu\mu}^{d,V} \approx 0.18$ are
  allowed by NuTeV. In fact, the two minima shown in
  Fig.~\ref{fig:NuTeV} are completely degenerate. }

For NSI with down-quarks the LMA-D degeneracy becomes disfavored at
more than 5$\sigma$, in agreement with the the results from
Ref.~\cite{Miranda:2004nb}; specifically for this case we have
$\Delta\chi^2_\text{OSC+SCAT,min}(\text{LMA-D)}=27$. However, our
results show that for NSI with up-quarks the LMA-D solution is still
allowed at 1.5$\sigma$
($\Delta\chi^2_\text{OSC+SCAT,min}(\text{LMA-D})=2.4$).  The
difference between the results for LMA-D for up and down quarks can be
easily understood as follows: LMA-D requires
$\eps^{q,V}_{ee}-\eps^{q,V}_{\mu\mu} \sim {\cal O}(-1)$ which given
the constraints on $\eps^{q,V}_{\mu\mu}$ for either $q=u$ or $d$
imposed by NuTeV (see Fig.~\ref{fig:NuTeV}), implies that
$\eps^{q,V}_{ee}\sim {\cal O}(-1)$, a value ruled out by CHARM for
$q=d$ but still allow for $q=u$ as seen in Fig.~\ref{fig:CHARM}.

\section{A future experiment on coherent neutrino-nucleus scattering}
\label{sec:COHERENT}

As mentioned in Sec.~\ref{sec:framework}, the bounds derived from
NuTeV and CHARM are not applicable to NSI models with light
mediators. In this case, it would be necessary to include constraints
from scattering experiments with low momentum transfer, in addition to
those from oscillations to constrain all the NSI
parameters. Neutrino-nucleus coherent scattering experiments can be
used for this purpose~\cite{Barranco:2005yy, Scholberg:2005qs,
  Lindner:2016wff, Dent:2016wcr}. Several proposals have been
envisaged for the future, which can be divided in two different
categories, according to their neutrino source: those using nuclear
reactors (\eg TEXONO~\cite{Wong:2008vk},
CONNIE~\cite{Aguilar-Arevalo:2016khx, Aguilar-Arevalo:2016qen},
MINER~\cite{Agnolet:2016zir}, or at the Chooz
reactor~\cite{Billard:2016giu}), and those using a stopped pion source
(COHERENT~\cite{Akimov:2015nza}). An important difference between the
two approaches is the flavor composition of the source.  A reactors
emits only electron anti-neutrinos and hence we can test only NSI
parameters $\eps_{e\alpha}$ ($\alpha = e,\mu,\tau$). The stopped pion
source provides neutrinos of muon and electron flavor which to some
extent can be disentangled by using timing information, as we explain
in more detail below. Hence, there we can constrain both,
$\eps_{e\alpha}$ and $\eps_{\mu\alpha}$.  To be specific, we will
consider in this work the COHERENT proposal~\cite{Akimov:2015nza} as
an example for a coherent neutrino-nucleon scattering experiment at a
stopped pion source, and comment on how the results would change in
case of a reactor measurement.  For recent sensitivity investigations
of a reactor based experiment see Refs.~\cite{Lindner:2016wff,
  Dent:2016wcr}.

COHERENT will place several low threshold detectors located within
tens of meters from the Spallation Neutrino Source
(SNS)~\cite{Kustom:2000rj} at Oak Ridge National Laboratory. A variety
of nuclear targets have been considered, including CsI, NaI, Ge, Ne
and Ar. The combination of different phases using different nuclear
targets would be beneficial not only because of the increase in
statistics, but also because it would allow to study the dependence of
the signal with $Z$ and $N$ (see Eq.~\eqref{eq:QwCOHERENT}). In this
work, for simplicity, we will only consider a \Ge{} detector, located
at a distance of 22~m from the source. We will assume a detection
threshold of 5~keV for nuclear recoils and a nominal exposure of 10
kg$\cdot$yrs, following Refs.~\cite{Akimov:2015nza, Cooper:2016}. In
order to illustrate the effect on the results of combining data taken
using different nuclei, for some of the results shown in
Sec.~\ref{sec:cohcomb} we will also add a second detector with a Ne
target.

At the SNS, the main component of the flux will be a monochromatic
$\nu_\mu$ line coming from $\pi^+ \to \mu^+ \, \nu_\mu$. Subleading
contributions to the flux come from the subsequent decays
$ \mu^+ \to e^+ \, \bar\nu_\mu \, \nu_e$    
(thus $|\vec p_\mu|=E_{\nu_\mu}=(m_\pi^2-m_\mu^2)/2/m_\pi = 30 $ MeV).
So the  the energy distributions (normalized to
1) for each neutrino flavor at the source can be obtained from simple
decay kinematics (neglecting the small $\mu$ momentum, ie taking the muon
at rest also), as
\begin{equation}
  \label{eq:COHflux}
  \begin{aligned}
    f_{\nu_\mu} &= \delta\left(E_\nu-\frac{m_\pi^2-m_\mu^2}{2m_\pi}\right) \,,
    \\
    f_{\bar\nu_\mu} &= \frac{64}{m_\mu}\left[\left(\frac{E_\nu}{m_\mu}\right)^2
      \left(\frac34-\frac{E_\nu}{m_\mu}\right)\right] \,,
    \\
    f_{\nu_e} &= \frac{192}{m_\mu}\left[\left(\frac{E_\nu}{m_\mu}\right)^2
      \left(\frac12-\frac{E_\nu}{m_\mu}\right)\right] \,,
  \end{aligned}
\end{equation}
where $E_\nu\in[0,m_\mu/2]$ is the energy of the resulting neutrino,
$m_\pi$ is the pion mass and $m_\mu$ is the muon mass.  As seen from
Eq.~\eqref{eq:COHflux}, the monochromatic $\nu_\mu$ flux line has an
energy of $\sim 30$~MeV, while the two other contributions will have a
continuous spectrum until they reach the end point of the decay at
around 50~MeV. The total flux $\phi (E_\alpha)$ is obtained
multiplying the distributions in Eq.~\eqref{eq:COHflux} by an overall
normalization factor, determined by the total number of protons on
target and the number of pions produced per incident proton. We set
this normalization constant following Ref.~\cite{Scholberg:2005qs}, so
that the total neutrino flux entering the detector is $10^7$ neutrinos
per second.

The coherent interaction cross section for a given neutrino flavor
$\alpha$, in presence of neutral-current NSI, can be written as
\begin{equation}
  \frac{d\sigma_\alpha}{dE_r} = \frac{G_F^2}{2\pi}
  \frac{Q_{w\alpha}^2}{4} F^2(2ME_r)M
  \left(2-\frac{ME_r}{E_\nu^2}\right) \,,
\end{equation}
where $E_r$ is the nuclear recoil energy, $F(Q^2)$ is the nuclear form
factor (taken from Ref.~\cite{Horowitz:2003cz}), $M$ is the mass of
the target nucleus and $E_\nu$ is the incident neutrino energy. We
have defined $Q_{\omega \alpha}^2$ as
\begin{multline}
  \frac{1}{4} Q_{w\alpha}^2 = \left[ Z(g^{V}_p+2\eps_{\alpha\alpha}^{u,V} +
    \eps_{\alpha\alpha}^{d,V})+N(g^{V}_n +\eps_{\alpha\alpha}^{u,V} +
    2\eps_{\alpha\alpha}^{d,V}) \right]^2
  \\
  + \sum_{\beta\neq\alpha}\left[Z(2\eps_{\alpha\beta}^{u,V} +
    \eps_{\alpha\beta}^{d,V})+N(\eps_{\alpha\beta}^{u,V} + 2\eps_{\alpha\beta}^{d,V})
    \right]^2 \,.
  \label{eq:QwCOHERENT}
\end{multline}
Here, $N$ and $Z$ are the number of neutrons and protons in the target
nucleus ($Z = 32$ and $N = 44$ for \Ge), respectively, while
$g_p^{V}=\frac12-2\sin^2\theta_W$, $g^{V}_n = - \frac12$ are the SM
couplings to the Z boson to protons and neutrons, $\theta_W$ being the
weak mixing angle.

The differential event distribution for a given flavor is obtained
from the convolution of the neutrino flux and cross section,
multiplying by appropriate normalization factors to account for the
total luminosity of the experiment. The result can be expressed as
\begin{equation}
  \label{eq:dNdE}
  \frac{dN_\alpha}{dE_r} =
  N_t \Delta t \int dE_\nu \phi_\alpha(E_\nu) \frac{d\sigma_\alpha}{dE_r}(E_\nu) \,,
\end{equation}
where $N_t$ is the number of nuclei in the detector, and $\Delta t$ is
the considered data taking period. In the absence of a detailed
publicly available simulation of the expected performance of the
COHERENT detector, we will assume perfect detection
efficiency\footnote{A lower detection efficiency can be easily
  corrected for by increasing the total exposure over the nominal
  value.} and will use no spectral information in our analysis (only
the total event rates, as explained below). The total number of events
is obtained after integrating Eq.~\eqref{eq:dNdE} over $E_r$ above
detection threshold.

We will apply a timing cut to separate the prompt signal (which comes
mainly from the monoenergetic $\nu_\mu$'s) from the delayed signal
(which comes mainly from the decay products of the muon). This
separation however is not perfect.  Considering that the muon lifetime
from a stopped pion is $\Gamma \tau=2.283$ $\mu$s, and that the SNS
proton pulses are relatively wide
($t_w=0.695~\mu$s)~\cite{Kustom:2000rj}, there is a certain
probability for a muon to decay within the duration of a given
pulse. In this case, the neutrinos produced from the muon decay may
contaminate the prompt signal window with a probability
\begin{equation}
  P_c = \frac1{t_w}\int_0^{t_w}dt[1-e^{-(t_w-t)/\Gamma\tau}] = 0.138 \,,
\end{equation}
where we have assumed a flat pulse shape. We have explictly checked that
the allowed values of NSI are hardly affected by this assumption or
even by modifications leading to changes of $P_c$ by factors
${\cal O}({\rm few})$.
The number of events detected within the prompt ($N_p$) and delayed
($N_d$) time windows are thus given by
\begin{equation}
  \begin{aligned}
    N_p & = N_{\nu_\mu}+P_c(N_{\nu_e}+N_{\bar\nu_\mu}) \,,\\
    N_d & = (1-P_c)(N_{\nu_e}+N_{\bar\nu_\mu})\,.
  \end{aligned}
\end{equation}
For the main configuration considered in this work, that is, a
\Ge~detector with a 5~keV threshold and a nominal exposure of
10~kg$\cdot$yr, we obtain approximately 113 (200) events in the prompt
(delayed) window. These numbers are in good agreement with
Ref.~\cite{Scholberg:2005qs}.

The experiment will be subject to systematic uncertainties affecting
the beam flux normalization, detector performance, etc. Following
Ref.~\cite{Scholberg:2005qs}, we estimate prior uncertainties to be at
the 10\% level. Significant backgrounds are expected from two main
sources: (1) beam-related backgrounds, especially fast neutrons which
enter the detector, and (2) backgrounds from cosmic ray interactions
and radioactivity. Based on Ref.~\cite{Akimov:2015nza}, we estimate
the number of background events to be approximately 20\% of the number
of signal events. We include them in our chi-square implementation
using the pull method, assuming they contribute to the statistical
error of the measurement.

Given that the expected statistics is in the range of a few hundred of
events per bin, a Gaussian $\chi^2$ is used for the COHERENT
experiment,
\begin{equation}
  \chi^2_\text{COH}=\min_{\xi} \sum_{k=p,d}
  \left( \frac{(1+\xi)N_{k,\text{NSI}}-N_{k,\text{obs}}}
       { \sqrt{N_{k,\text{obs}} + 0.2 N_{k,\text{obs}}} } \right)^2 +
  \left( \frac{\xi }{\sigma_{sys} } \right)^2 \,,
  \label{eq:chi2coh}
\end{equation}
where $ \sigma_{sys} = 0.1$ as explained above, and the result is
minimized over the nuisance parameter $\xi$ associated to the signal
normalization. Here, $N_{p,\text{obs}}$ and $N_{d,\text{obs}}$ denote
the simulated data that we assume the experiment will observe in the
prompt and delayed time windows, respectively, while the corresponding
expected values in presence of NSI parameters are denoted as
$N_{p,\text{NSI}}$ and $N_{d,\text{NSI}}$.

\begin{figure}\centering
  \includegraphics[width=0.45\textwidth]{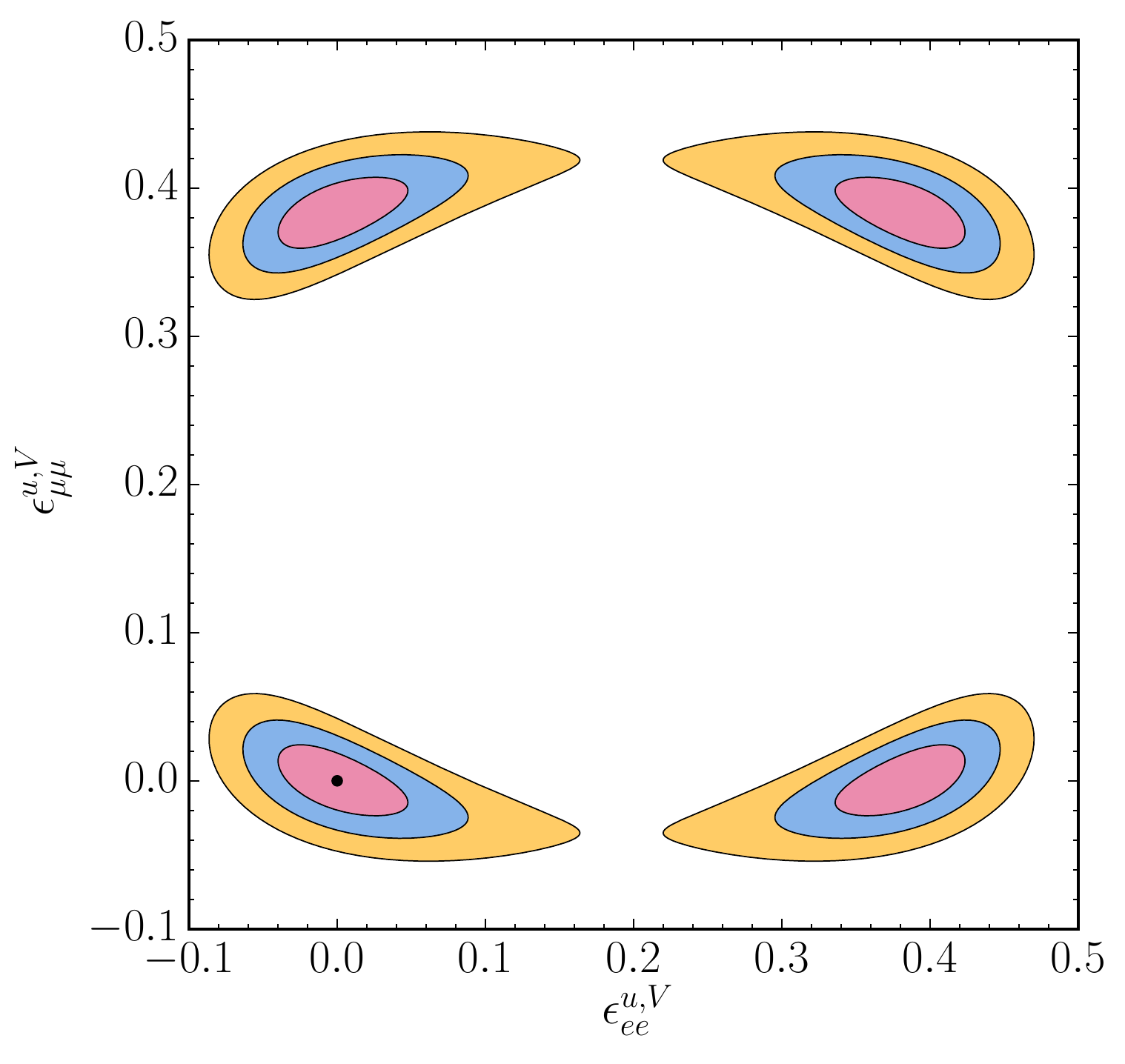}
  \quad
  \includegraphics[width=0.45\textwidth]{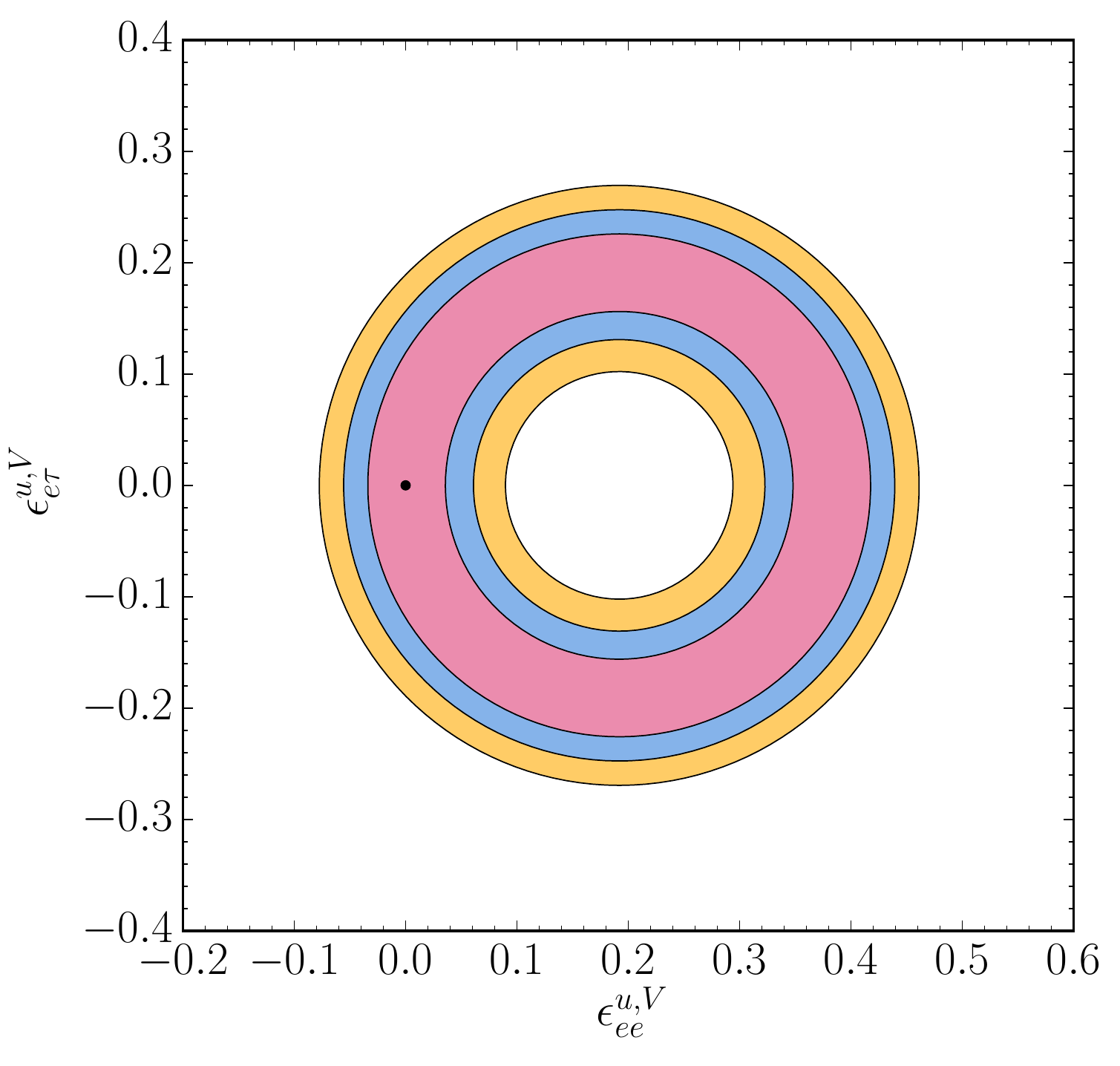}
  \caption{The projected COHERENT sensitivity to NC vector-like NSI
    parameters. The left (right) panel shows the expected confidence
    regions projected onto the $\epsilon^{u,V}_{ee} -
    \epsilon^{u,V}_{\mu\mu}$ ($\epsilon^{u,V}_{ee} -
    \epsilon^{u,V}_{e\tau} $) plane, after setting the remaining NSI
    parameters to zero. In both cases, the pink, blue and yellow
    regions indicate the allowed regions at 1$\sigma$, 2$\sigma$ and
    3$\sigma$ (for 2 d.o.f.).  The black dots indicate the SM, which
    has been used to generate the simulated experimental data used in
    this figure. The experiment has been simulated using
    10~kg$\cdot$yr exposure for \Ge, see Sec.~\ref{sec:COHERENT} for
    details.}
  \label{fig:COHERENT-sensitivity}
\end{figure}

Fig.~\ref{fig:COHERENT-sensitivity} shows the expected sensitivity
for the COHERENT setup, simulated as described above, to several NSI
parameters affecting neutrino interactions with up quarks (the
corresponding regions for interactions with down quarks are very
similar). In both panels, SM interactions (\ie zero NSI) have been
assumed to simulate the COHERENT ``observed'' data, and the result is
fitted allowing for the presence of NSI. All NSI parameters not shown
in each panel have been set to zero for simplicity in this figure; in
the results shown in the next section, however, they are all included
and the chi-squared is minimized over all parameters not shown.

A coherent scattering experiment at a reactor would be sensitive only
to the NSI combination $Q_{we}^2$, as defined in
Eq.~\eqref{eq:QwCOHERENT}. Hence we would obtain a qualitatively
similar behavior of the NSI sector involving the electron flavor (for
instance as shown in the right panel of
Fig.~\ref{fig:COHERENT-sensitivity}). In contrast to the configuration
shown in the left panel, there would be no sensitivity to
$\eps_{\mu\mu}^{q,V}$ from a reactor experiment. This will turn out to
be crucial for resolving the LMA-D degeneracy, as we will discuss
below.

\section{Expected combined sensitivity after inclusion of COHERENT}
\label{sec:cohcomb}

In this section we add to our global fit the expected results for the
COHERENT experiment. As before, we consider two scenarios depending on
the assumed mass range of the mediator responsible for the NSI. In all
cases, we will show our results as $\Delta \chi^2 \equiv \chi^2 -
\chi^2_\text{min}$, where $\chi^2_\text{min}$ is the global minimum of
the $\chi^2$. The particular combination of experiments included in
the $\chi^2$ will depend on the scenario being considered, as
described in more detail below. As before, we will consider real NSI
and will assume vanishing axial NSI couplings.

\subsection{NSI from a light mediator}

As explained in Sec.~\ref{sec:framework}, if NSI are produced from new
interactions with a light mediator the only applicable bounds are
those obtained from experiments with small momentum transfer, \ie
those derived from oscillation data and COHERENT. Therefore in this
case we construct our combined chi-squared function as
\begin{equation}
  \chi^2_\text{light,future}=\chi^2_\text{OSC}+\chi^2_\text{COH} \,.
\end{equation}
In the case of oscillations, we will use the results from the global
fit performed in Ref.~\cite{Gonzalez-Garcia:2013usa}. In the case of
COHERENT, some assumption needs to be made regarding the ``true''
values of the NSI parameters which will be used as input to generate
the simulated data. A natural possibility would be to set all $\eps$
to zero. A second possibility would be to use the best-fit from
oscillation data, which shows a slight preference for non-zero NSI. We
will consider those two cases below.

\begin{figure}\centering
  \includegraphics[width=\textwidth]{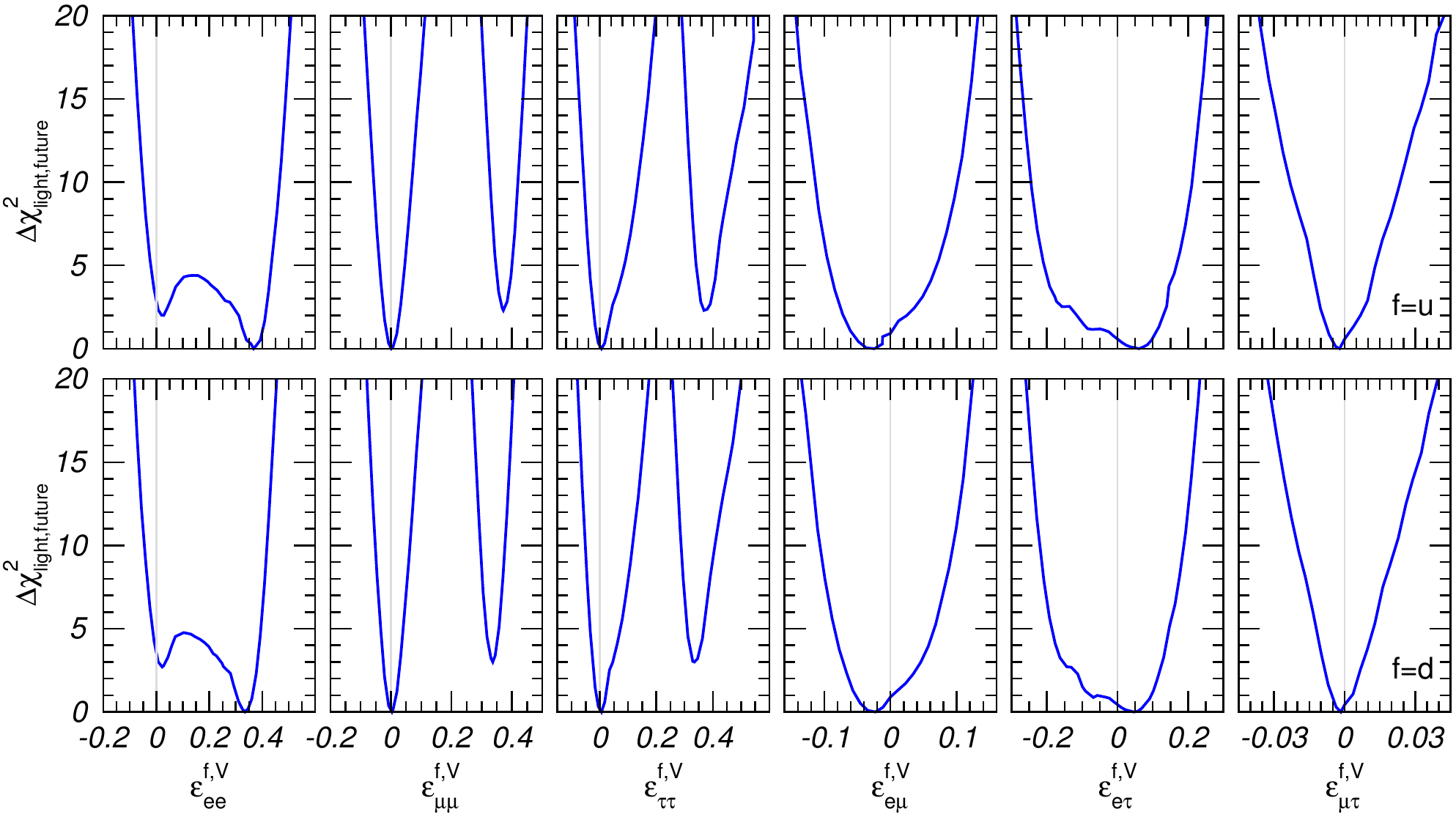}
  \caption{Dependence of $\Delta\chi^2_\text{light,future}$ on the NSI
    parameters $\eps_{\alpha\beta}^{f,V}$ for $f=u$ (upper panels) and
    $f=d$ (lower panels), using current global data on oscillations
    combined with artificial COHERENT data generated for all
    $\epsilon_{\alpha\beta}$ set to zero.}
  \label{fig:1Dlight}
\end{figure}

\begin{figure}\centering
  \includegraphics[width=0.9\textwidth]{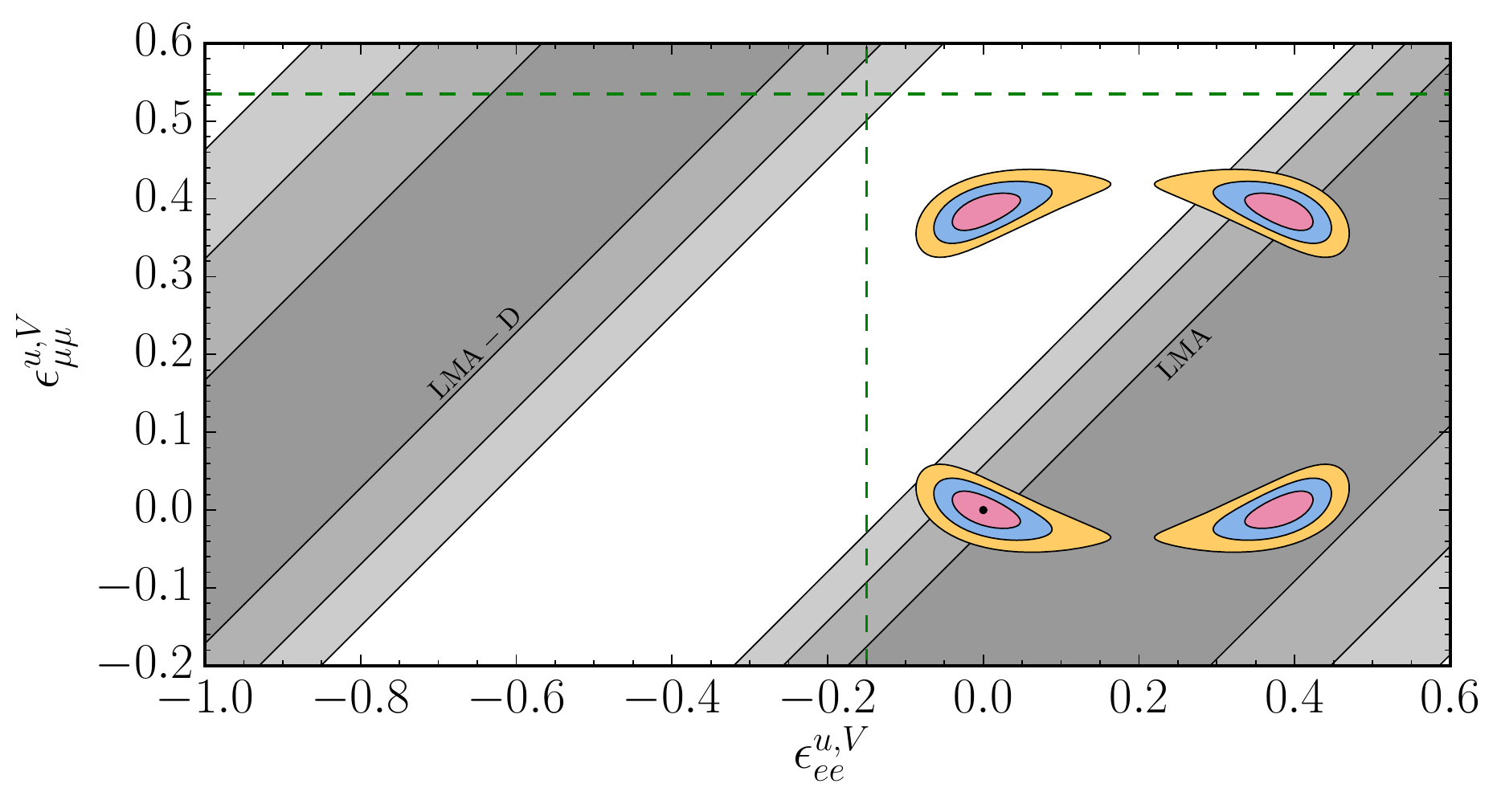}
  \caption{Allowed regions in the plane of $\eps^{u,V}_{ee}$ and
    $\eps^{u,V}_{\mu\mu}$ from the COHERENT experiment (under the
    assumption of no NSI in the data --- same as in
    Fig.~\ref{fig:COHERENT-sensitivity}) overlayed with the presently
    allowed regions from the global oscillation analysis. The two
    diagonal shaded bands correspond to the LMA and LMA-D regions as
    indicated, at 1, 2, 3$\sigma$. The dashed lines indicate the
    values of NSI parameters for which COHERENT would not be able to
    resolve the LMA-D degeneracy, see appendix~\ref{sec:app} for
    details.}
  \label{fig:2Dlight}
\end{figure}

Fig.~\ref{fig:1Dlight} shows the results for the combination of
oscillation data, plus COHERENT data simulated for vanishing NSI
coefficients. In generating these results we have assumed our template
COHERENT configuration of 10 kg$\cdot$yrs of \Ge{} with a threshold of
5 keV for nuclear recoils, see Sec.~\ref{sec:COHERENT} for
details. The results are shown for the NSI coefficients assuming that
the new interactions take place with either up or down quarks, as
indicated by the labels in each row. As mentioned earlier, we include
all NSI parameters at once in the fit. Thus, in each panel, the
results have been obtained after marginalization over all parameters
not shown, including standard oscillation parameters and NSI
parameters. For this configuration we find that the LMA-D solution can
be ruled out (for NSI with up or with down quarks) at high CL. In
particular we obtain $\Delta\chi^2_\text{light,future}(\text{LMA-D})>
45\,(80)$ for NSI with up (down) quarks. This is obvious from
Fig.~\ref{fig:2Dlight}, where we show the allowed regions in the plane
of $\eps^{u,V}_{ee}$ and $\eps^{u,V}_{\mu\mu}$ from oscillations
together with the 4 degenerate solutions from COHERENT (same as in
Fig.~\ref{fig:COHERENT-sensitivity}). The regions from oscillations
are diagonal bands in this plane, since oscillations determine only
the difference $\eps^{u,V}_{ee}-\eps^{u,V}_{\mu\mu}$. We see that the
band corresponding to the LMA-D region is far away from the COHERENT
solutions and can therefore be excluded by the combination.
Consequently, in Fig.~\ref{fig:1Dlight} only the results obtained for
the LMA solution appear. The corresponding allowed ranges at 90\% CL
are reported in Table~\ref{tab:90CL}.

For the LMA solution, comparing to the present bounds from
oscillations (see Fig.~\ref{fig:1Dpresent}), we see that no
significant improvement is expected in the determination of the
flavor-changing NSI parameters.  The main impact of COHERENT is in the
determination of the flavor diagonal ones: as it provides information
on $\eps^{q,V}_{ee}$ and $\eps^{q,V}_{\mu\mu}$, the combination with
oscillations allows for the independent determination of the three
flavor-diagonal couplings. However, three minima still remain for the
combined chi-squared, one global and two quasi-degenerate local. This
is explained as follows.  First, COHERENT is completely insensitive to
$\epsilon_{\tau\tau}^{f,V}$, as shown in
Eq.~\eqref{eq:QwCOHERENT}. This means that $\eps^{f,V}_{\mu\mu}$ can
be different from zero, as long as $\eps^{f,V}_{\tau\tau}$ is set
accordingly in order to respect the bounds from oscillations, which
constrain $\eps^{f,V}_{\tau\tau}-\eps^{f,V}_{\mu\mu} \approx 0 $.
Second, the shape of $\chi^2_\text{COH}$, as shown in
Fig.~\ref{fig:COHERENT-sensitivity}, has four separate minima in the
plane of $\eps^{f,V}_{ee}$ and $\eps^{f,V}_{\mu\mu}$.  As can be seen
from Fig.~\ref{fig:2Dlight}, the position of the low right region
matches very well the allowed values from oscillation data for the LMA
case $\eps^{f,V}_{ee} - \eps^{f,V}_{\mu\mu} \approx 0.3$.  This leads
to the global minimum observed in the one dimensional projection of
the combined $\Delta\chi^2_\text{light,,future}$ for $\eps_{ee}^{f,V}
\approx 0.35$ and $\eps_{\mu\mu}^{f,V}\simeq\eps_{\tau\tau}^{f,V}
\approx 0$, see Fig.~\ref{fig:1Dlight}.  In addition, there are two
other sets of NSI for which local minima are found in the
multi-dimensional parameter space. In the $(\eps^{f,V}_{ee},
\eps^{f,V}_{\mu\mu})$ plane shown in Fig.~\ref{fig:2Dlight} they
correspond to the lower left and upper right COHERENT regions.
Combined with the oscillation constraint
$\eps^{f,V}_{\tau\tau}-\eps^{f,V}_{\mu\mu} \approx 0$, the first set
corresponds to $\eps^{f,V}_{ee} \simeq \eps^{f,V}_{\mu\mu}
\simeq\eps^{f,V}_{\tau\tau}=0$, which is visible as the local minimum
in the first panel in Fig.~\ref{fig:1Dlight}. The second set
corresponds to $\eps^{f,V}_{ee} \simeq \eps^{f,V}_{\mu\mu}
\simeq\eps^{f,V}_{\tau\tau}=0.3$ which appears as the local minimum in
the second and third panels of Fig.~\ref{fig:1Dlight}.  In both cases
the oscillation probability has no flavor-diagonal NSI effects and
yields about the same $\Delta\chi^2_\text{OSC}\simeq 2$ (see first
panel in Fig.~\ref{fig:nsi-chisq}).

\bigskip

Let us now relax the assumption that the true values of the NSI
parameters are zero. To generate COHERENT data we adopt now the best
fit point obtained in the oscillation analysis for light mediators,
see Sec.~\ref{ssec:osc}. However, since oscillations are sensitive
only to the differences of flavor-diagonal NSI, one of the
$\eps_{\alpha\alpha}^{q,V}$ remains undetermined and can be chosen
arbitrarily. We use $\eps^{q,V}_{ee,\text{true}}$ as independent
diagonal parameter. We can now perform the combined
oscillation+COHERENT fit, by scanning the value of
$\eps^{q,V}_{ee,\text{true}}$, and all other NSI parameters assumed to
generate COHERENT data are determined by the best fit point from
oscillations (in particular, also the other two diagonal parameters
$\eps^{q,V}_{\mu\mu,\text{true}}$ and $\eps^{q,V}_{\tau\tau,\text{true}}$).

\begin{figure}\centering
  \includegraphics[width=0.9\textwidth]{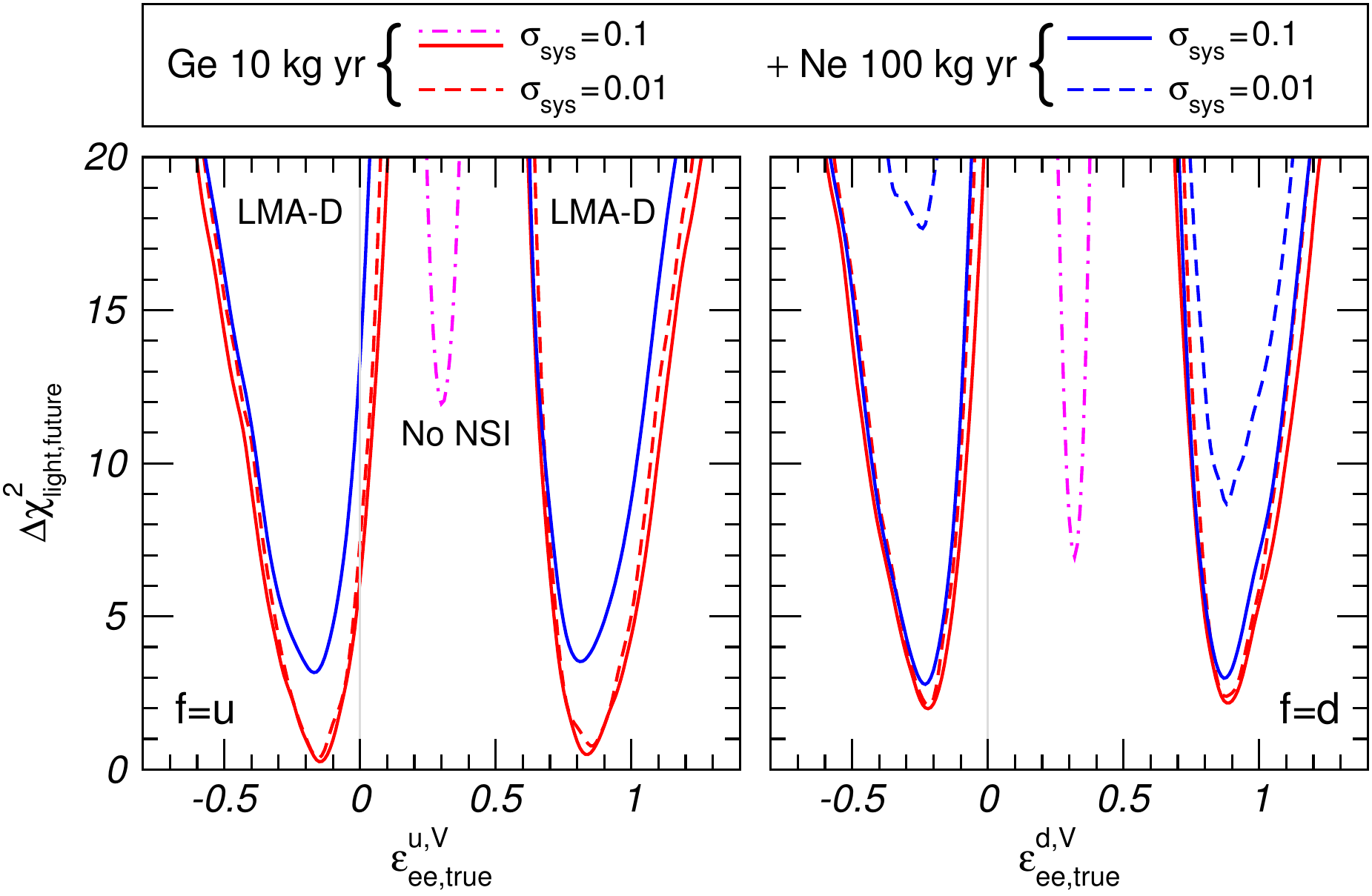}
  \caption{The full and dashed lines show
    $\Delta\chi^2_\text{min}(\text{LMA-D})$ from a combination of
    current oscillation and future COHERENT data (for different
    assumptions on target and systematics as labeled) as a function of
    the assumed value of $\eps^{q,V}_{ee,\text{true}}$. All other true
    values used to generate COHERENT data are set to the current best
    fit point from oscillations. To determine the minimum of the joint
    $\chi^2$ in the LMA-D region all oscillation and NSI parameters
    are varied. The dash-dotted curve shows the $\Delta\chi^2$ for
    no-NSI under the same assumptions for our default COHERENT
    configuration.}
  \label{fig:LMA-D-light-future}
\end{figure}

Let us focus on the question of whether the LMA-D degeneracy can be
lifted by the combination of oscillation and COHERENT data if
$\eps^{q,V}_{ee,\text{true}}$ is allowed to take on arbitrary
values. The full red curve in Fig.~\ref{fig:LMA-D-light-future} shows
the $\Delta\chi^2_\text{min}$ of the LMA-D region as a function of
$\eps^{q,V}_{ee,\text{true}}$ for our default COHERENT configuration
(\Ge{} detector with 5~keV threshold, 10 kg$\cdot$yrs, and 10\%
normalization systematics).  We find that there are two local minima
of this curve, which means that there are certain values of
$\eps^{q,V}_{ee,\text{true}}$ for which the combination of oscillation
and COHERENT data will not be able to resolve the LMA-D
degeneracy. The location of the minima can be understood analytically
from the combinations of NSI parameters which COHERENT is sensitive to
according to Eq.~\eqref{eq:QwCOHERENT}. We provide the relevant
equations in appendix~\ref{sec:app}. The values of $\eps^{q,V}_{ee}$
and $\eps^{q,V}_{\mu\mu}$ for which COHERENT cannot resolve the
degeneracy are shown as dashed lines in Fig.~\ref{fig:2Dlight}. The
region where they cross the allowed band from LMA corresponds to the
location of the minima in Fig.~\ref{fig:LMA-D-light-future}. In those
locations COHERENT is completely blind to the degeneracy and the
$\chi^2$ seen in the figure is just the one present already in the
oscillation-only analysis. Moreover, the figure shows that there is a
relevant region of the parameter space around those minima, where the
LMA-D degeneracy would remain at low CL. Also as seen in the figure,
reducing the normalization systematics in COHERENT to 1\% (dashed red
line) has a negligible impact.

However, if Nature happens to chose parameter values close to those
points, the oscillation+COHERENT combination will be able to establish
the existence of NSI at very high confidence. This is shown by the
dot-dashed curve in Fig.~\ref{fig:LMA-D-light-future}, which gives the
$\Delta\chi^2$ for the case with all $\eps_{\alpha\beta}^{q,V} = 0$.
This curve does not reach zero because (a) oscillation data show some
preference for non-zero NSI and give a contribution of $\Delta\chi^2 =
5.4$, see Sec.~\ref{ssec:osc}, and (b) for any value of the assumed
$\eps^{q,V}_{ee,\text{true}}$ and assuming the oscillation best-fit to
generate COHERENT data, the no-NSI point is disfavored at some level
by COHERENT. But what is clear from the figure is that the no-NSI
curve has no overlap with the regions where the LMA-D solution is a
problem for COHERENT+oscillations. Hence, we conclude that if
NSI exist with values such the LMA-D degeneracy remains, the
combination of the present oscillation results with those from our
default COHERENT set-up will tell us with high CL that non-zero NSI
are present.

The same conclusion can be drawn from Fig.~\ref{fig:2Dlight} for
general values of $\eps^{q,V}_{ee}$ and $\eps^{q,V}_{\mu\mu}$,
independent of the current best-fit point, by noting that the two
dashed lines (along which the LMA-D degeneracy cannot be resolved) are
``far'' from the COHERENT solutions in case of no NSI and hence,
parameter values along those lines can be distinguished from no NSI
with high significance.

In principle the blind spots of COHERENT to the degeneracy shown in
Fig.~\ref{fig:LMA-D-light-future} could be lifted by using multiple
nuclear targets since, as shown in appendix~\ref{sec:app}, the
locations of the blind spots depend on $Z$ and $N$ for the used
nucleon. However, quantitatively the effect is rather small and adding
Ar or Ne to Ge within our standard setup of COHERENT described in
Sec.~\ref{sec:COHERENT} does not lead to much change. This is
illustrated by the full blue line in
Fig.~\ref{fig:LMA-D-light-future}. This line shows the
$\Delta\chi^2_\text{min}$ of the LMA-D region after adding data
corresponding to 100 kg$\cdot$yr exposure for $^{20}$Ne (with a more
conservative 10~keV threshold)\footnote{The choice of $^{20}$Ne is not
  arbitrary. Among the considered targets for the COHERENT experiment,
  it gives the most different Z/N ratio compared to \Ge. This will
  provide the largest effect on the combined sensitivity. } on top of
the results obtained for our default \Ge~configuration. We find that
this combination can partially lift the degeneracy, but only when the
systematic normalization error for each of these data samples is
substantially reduced from our default 10\% value.  This is shown by
the dashed blue curve in the figure where we show the results for the
combination of \Ge{} and $^{20}$Ne but reducing the systematic
uncertainty down to a 1\% systematic normalization uncertainty (which
is taken to be completely uncorrelated between the two contributions
to the COHERENT total $\chi^2$).

Let us finally comment on the expected changes on these results if a
coherent-scattering data from a reactor based experiment is used
instead of the stopped pion source setup. In that case only the
combination $Q_{we}^2$ is determined, providing no constraint on
$\eps_{\mu\mu}^{q,V}$. In this case, the degeneracy can be shifted
completely in the $\mu\mu$ and $\tau\tau$ sector, and it will not be
possible to lift the LMA-D degeneracy, for any true value of
$\eps_{ee}^{q,V}$, see also appendix~\ref{sec:app}. The situation is
different in the heavy mediator case, where $\eps_{\mu\mu}^{q,V}$ is
strongly constrained by NuTeV, as we will see below. In this case a
coherent-scattering measurement at a reactor setup should suffice to
rule out the LMA-D degeneracy.

\subsection{NSI from a heavy mediator}

Next we consider the case of NSI induced in models with a heavy
mediator, as introduced in Sec.~\ref{ssec:prescombheavy}.  In this
scenario, the scattering bounds from deep-inelastic neutrino-nucleus
scattering would also apply. Thus, we construct a combined statistics
including bounds from oscillations, CHARM, NuTeV, and the expected
future contribution from COHERENT, as
\begin{equation}
  \chi^2_\text{heavy,future}=
  \chi^2_\text{OSC}+\chi^2_\text{CHARM}+\chi^2_\text{NuTeV}+\chi^2_\text{COH} \,.
\end{equation}
Again in this case, to simulate the COHERENT data some assumption
needs to be made regarding the input values of the NSI parameters.
However, for this scenario the addition of NuTeV and CHARM to the
oscillation data already provides very strong constraints on the NSI
parameters (see Fig.~\ref{fig:1Dpresent}) and, thus, the results
obtained simulating COHERENT data for vanishing NSI, or for NSI
according to the best-fit values of oscillation plus scattering, are
very similar. Therefore, in this section we will only consider the
case when COHERENT data are simulated using the SM as input (\ie all
NSI coefficients set to zero). A COHERENT configuration of 10
kg$\cdot$yrs of \Ge, with a threshold of 5 keV for detection of
nuclear recoils, will be considered as explained in more detail in
Sec.~\ref{sec:COHERENT}.

\begin{figure}\centering
  \includegraphics[width=\textwidth]{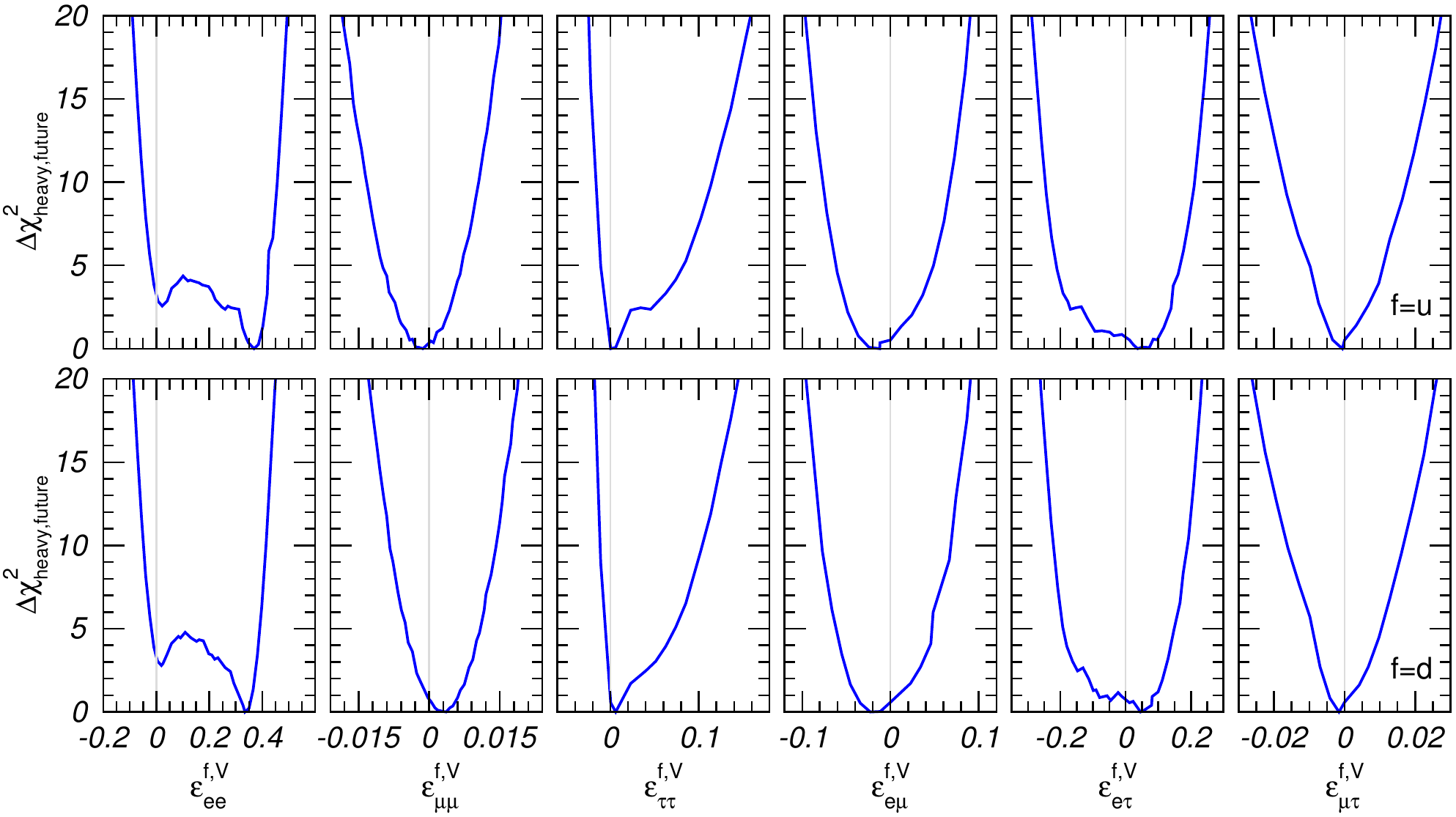}
  \caption{Dependence of $\Delta\chi^2_\text{heavy,future}$ on the NSI
    parameters $\eps_{\alpha\beta}^{f,V}$ for $f=u$ (upper panels) and
    $f=d$ (lower panels), using current global data on oscillations
    and DIS data from NuTeV and CHARM combined with artificial
    COHERENT data generated for all $\epsilon_{\alpha\beta}$ set to
    zero.}
  \label{fig:1Dheavy}
\end{figure}

Fig.~\ref{fig:1Dheavy} shows the dependence of
$\Delta\chi^2_\text{heavy,future}$ with all NSI coefficients (after
marginalizing over all oscillation and NSI undisplayed parameters),
for interactions with either up or down quarks as indicated by the
labels. The corresponding allowed ranges at 90\% CL are reported in
Table~\ref{tab:90CL}. As can be seen from the comparison between
Fig.~\ref{fig:1Dheavy} and~\ref{fig:1Dpresent}, the addition of
COHERENT data allows to reject the LMA-D solution also in the $f=u$
scenario at high confidence level
($\Delta\chi^2_\text{heavy,future,min}(\text{LMA-D})\sim 100$).  For
the LMA solution, the assumed configuration of COHERENT improves the
determination of $\eps^{f,V}_{ee}$, while for all other NSI parameters
we find no significant improvement compared to the present analysis of
oscillations and scattering experiments.

\section{Summary and conclusions}
\label{sec:conclusions}

\begin{table}\centering
  \catcode`!=\active \def!{\hphantom{$-$}}
  \catcode`?=\active \def?{\hphantom{0}}
  \caption{90\% and allowed ranges for the NSI parameters
    $\eps_{\alpha\beta}^f$ for $f=u,d$ as obtained from the different
    combined analyses.  The upper (lower) part of the table
    corresponds to models of NSI's generated by light (heavy)
    mediators. The results in each panel are obtained after
    marginalizing over oscillation and the other NSI parameters.  See
    text for details.}
  \label{tab:90CL}
  \begin{tabular}{|c|c|c|c|}
    \hline
    \multicolumn{4}{|c|}{Light}
    \\
    \hline
    \multicolumn{2}{|c|}{PRESENT (OSC)}
    & \multicolumn{2}{c|}{+COHERENT(SM)}
    \\
    \hline
    \multirow{2}{*}{$\eps_{ee}^{u,V}-\eps_{\mu\mu}^{u,V}$}
    & \multirow{2}{*}{$[-1.19, -0.81] \oplus [0.00, 0.51] $}
    & $\eps_{ee}^{u,V}$
    & !$[0.002, 0.049] \oplus [0.28, 0.42]$
    \\[1mm]
    && $\eps_{\mu\mu}^{u,V}$
    & $[-0.026, 0.033] \oplus [0.36, 0.38]$
    \\[-2mm]
    $\eps_{\tau\tau}^{u,V}-\eps_{\mu\mu}^{u,V}$
    & $[-0.03, 0.03]$
    &&
    \\[-2mm]
    && $\eps_{\tau\tau}^{u,V}$
    & $[-0.025, 0.047] \oplus [0.36, 0.39]$
    \\[+3mm]
    $\eps_{e\mu}^{u,V}$
    & $[-0.09, 0.10]$
    & \multicolumn{2}{c|}{$[-0.08,0.04]$}
    \\
    $\eps_{e\tau}^{u,V}$
    & $[-0.15, 0.14]$
    & \multicolumn{2}{c|}{$[-0.17, 0.14]$}
    \\
    $\eps_{\mu\tau}^{u,V}$
    & $[-0.01, 0.01]$
    & \multicolumn{2}{c|}{$[-0.01, 0.01]$}
    \\
    \hline
    \multirow{2}{*}{$\eps_{ee}^{d,V} - \eps_{\mu\mu}^{d,V}$}
    & \multirow{2}{*}{$ [-1.17, -1.03] \oplus [0.02, 0.51]$}
    & $\eps_{ee}^{d,V}$
    & $[0.022, 0.023] \oplus [0.25, 0.38]$
    \\[+1mm]
    && $\eps_{\mu\mu}^{d,V}$
    & $[-0.024, 0.029]$
    \\[-2mm]
    $\eps_{\tau\tau}^{d,V}-\eps_{\mu\mu}^{d,V}$
    & $[-0.01, 0.03]$
    &&
    \\[-2mm]
    && $\eps_{\tau\tau}^{d,V}$
    & $[-0.023, 0.039]$
    \\[+3mm]
    $\eps_{e\mu}^{d,V}$
    & $[-0.09, 0.08]$
    & \multicolumn{2}{c|}{$[-0.07, 0.04]$}
    \\
    $\eps_{e\tau}^{d,V}$
    & $[-0.13, 0.14]$
    & \multicolumn{2}{c|}{$[-0.14, 0.12]$}
    \\
    $\eps_{\mu\tau}^{d,V}$
    & $[-0.01, 0.01]$
    & \multicolumn{2}{c|}{$[-0.009, 0.007]$}
    \\
    \hline
    \hline
    \multicolumn{4}{|c|}{Heavy}
    \\
    \hline
    \multicolumn{2}{|c|}{PRESENT (OSC+CHARM+NuTeV)}
    & \multicolumn{2}{c|}{+COHERENT(SM)}
    \\
    \hline
    $\eps_{ee}^{u,V}$
    & $[-0.97, -0.83] \oplus [0.033,0.450]$
    & \multicolumn{2}{c|}{$[0.014, 0.032] \oplus [0.24, 0.41]$}
    \\
    $\eps_{\mu\mu}^{u,V}$
    & $[-0.008, 0.005]$
    & \multicolumn{2}{c|} {$[-0.007, 0.005]$}
    \\
    $\eps_{\tau\tau}^{u,V}$
    & $[-0.015, 0.04]$?
    & \multicolumn{2}{c|}{$[-0.006, 0.04]$?}
    \\
    $\eps_{e\mu}^{u,V}$
    & $[-0.05, 0.03]$
    & \multicolumn{2}{c|}{$[-0.05, 0.03]$}
    \\
    $\eps_{e\tau}^{u,V}$
    & $[-0.15, 0.13]$
    & \multicolumn{2}{c|}{$[-0.15, 0.13]$}
    \\
    $\eps_{\mu\tau}^{u,V}$
    & $[-0.006,0.005]$
    & \multicolumn{2}{c|}{$[-0.006, 0.004]$}
    \\
    \hline
    $\eps_{ee}^{d,V}$
    & !$[0.02, 0.51]$
    & \multicolumn{2}{c|}{!$[0.26, 0.38]$}
    \\
    $\eps_{\mu\mu}^{d,V}$
    & $[-0.003, 0.009]$
    &\multicolumn{2}{c|} {$[-0.003, 0.009]$}
    \\
    $\eps_{\tau\tau}^{d,V}$
    & $[-0.001, 0.05]$?
    & \multicolumn{2}{c|}{$[-0.001, 0.05]$?}
    \\
    $\eps_{e\mu}^{d,V}$
    & $[-0.05, 0.03]$
    & \multicolumn{2}{c|}{$[-0.05, 0.03]$}
    \\
    $\eps_{e\tau}^{d,V}$
    & $[-0.15, 0.14]$
    & \multicolumn{2}{c|}{$[-0.15, 0.14]$}
    \\
    $\eps_{\mu\tau}^{d,V}$
    & $[-0.007, 0.007]$
    & \multicolumn{2}{c|}{$[-0.007,0.007]$}
    \\
    \hline
  \end{tabular}
\end{table}

Non-Standard neutrino interactions (NSI) are generic expectations of
physics beyond the standard model and can be parametrized in a
model-independent approach in terms of dimension-six operators, which
arise as the low-energy limit of some new interaction after
integrating out its mediator.  NSI modifying the charged-current
leptonic interactions are currently strongly constrained by charged
lepton data, while data on NSI affecting the neutral-current
interactions (NSI-NC) of the neutrinos are sparse.  At present current
global fits to oscillation data provide some of the strongest
constraints on NSI-NC, in particular for vector-like interactions
which are those which affect the flavor evolution of the neutrinos in
matter.  Still, the results obtained in
Ref.~\cite{Gonzalez-Garcia:2013usa} show that there remain two sets of
solutions compatible with the data: the so-called LMA solution, as
given in the SM extended with neutrino masses and mixing, compatible
with negligible NSI, and a second one dubbed
LMA-Dark~\cite{Miranda:2004nb} (or LMA-D), which requires large NSI
and the solar mixing angle to lie in the upper octant. Currently, the
two solutions (LMA and LMA-D) are almost completely degenerate, the
fit showing only a slight preference for the LMA solution with $\Delta
\chi^2 \sim 0.2 (2)$ for the NSI with up (down) quarks.

The LMA-D solution is a consequence of a more profound degeneracy
which affects the Hamiltonian governing neutrino oscillations. This
degeneracy involves a change in the matter potential and a change in
the octant of the solar angle, but it also needs a change in the
CP-phase $\delta$ and a flip in the neutrino mass ordering. Besides,
it takes place regardless of whether the experiment is performed in
vacuum or in presence of a matter potential. Therefore, this
degeneracy will make it impossible to determine the neutrino mass
ordering at neutrino oscillation experiments unless it is ruled out by
other experiments first.

Other processes capable of constraining NSI-NC include
neutrino-nucleus deep-inelastic scattering. At present, the most
precise results come from NuTeV and CHARM and provide constraints for
NSI affecting $\nu_\mu$ and $\nu_e$ respectively. Since oscillation
data is only sensitive to differences among diagonal couplings, the
combination with scattering data is crucial to obtain independent
bounds for all parameters entering the matter potential separately. In
this work, we have performed a global fit to oscillation and
scattering data from the NuTeV and CHARM experiments, deriving the
strongest constraints on neutral-current NSI in the literature. The
fit was done including all vector NSI operators affecting either up or
down quarks at a time. Marginalization over the standard oscillation
parameters has been performed. Our results for this fit are summarized
in Fig.~\ref{fig:1Dpresent}, while the limits at the 90\% CL are
listed in Table~\ref{tab:90CL}.  The combination with scattering data
also rules out the LMA-D solution for NSI involving down quarks, as
pointed out earlier~\cite{Miranda:2004nb, Escrihuela:2009up}. However,
we show that the LMA-D solution still survives for NSI with up quarks.

Nevertheless, as we have stressed, the NuTeV and CHARM bounds are not
applicable to all models leading to NSI in the neutrino sector.  For
example, if the NSI come from neutrino interactions with a new light
mediator, the bounds derived from deep-inelastic processes will be
strongly suppressed with the inverse of the momentum transfer and can
be evaded. In this case, only oscillation data would be
applicable. Future neutrino-nucleus coherent scattering experiments
will also be able to put additional constraints. As coherent
neutrino-nucleus scattering involves a much lower momentum transfer,
such bounds would be applicable in models with light mediators. Thus,
in the second part of our work, we have explored the impact of the
results expected from such experiments on the limits to NSI operators
taking as an example the COHERENT experiment with a \Ge{} detector
with 5~keV threshold and 10 kg$\cdot$yrs exposure at a stopped pion
source.  We have distinguished explicitly two cases: NSI models with
heavy mediators (where bounds from oscillation data, NuTeV and CHARM
would apply) and models with light mediators (where only present
oscillation bounds would apply). In order to generate COHERENT data,
we have used two assumptions: (i) the data are obtained under the
assumption of no NSI, and (ii) the data are obtained using the
best-fit NSI values from a global fit to previous experiments. Our
results are summarized in Figs.~\ref{fig:1Dheavy} for the heavy
mediator case, and in Figs.~\ref{fig:1Dlight}
and~\ref{fig:LMA-D-light-future} for the light mediator case. The
expected 90\% CL ranges are summarized in Table~\ref{tab:90CL}.

In the case of NSI from light mediators, we find that the combination
of COHERENT and current experiments should be able to definitely rule
out the LMA-D solution also in the case of NSI with up quarks, as long
as the results of COHERENT are as expected for negligible NSI (case i
above). However, if COHERENT data is instead in agreement with the
expectations from the current best-fit point of oscillations it may
not be possible to rule out the LMA-D solution, as clearly illustrated
in Fig.~\ref{fig:LMA-D-light-future}. This is a consequence of the
presence of degeneracies in the NSI parameters allowed by COHERENT, as
we detail in appendix~\ref{sec:app}.  We find that breaking those and
fully ruling out the LMA-D may be achieved by a combination of
coherent scattering data with different nuclei, but only if very good
control of the systematics affecting the normalization of the event
rates can be achieved.  However, one must realize that, even in
scenarios for which the LMA-D degeneracy cannot be lifted, the
experiment will be able to rule out the no-NSI hypothesis at high CL
and discover the presence of new interactions in the neutrino sector.

Conversely, in the heavy mediator case, when CHARM and NuTeV
constraints apply, adding COHERENT data will rule out the LMA-D region
also for the case of NSI with up quarks. Hence, in this case it is always
possible to completely resolve the degeneracy.

\acknowledgments

We warmly thank Kate Scholberg for useful discussions and for
providing us with the form factors needed to simulate the COHERENT
experiment. This work is supported by USA-NSF grant PHY-1620628, by EU
Networks FP10 ITN ELUSIVES (H2020-MSCA-ITN-2015-674896) and
INVISIBLES-PLUS (H2020-MSCA-RISE-2015-690575), by MINECO grant
FPA2013-46570 and MINECO/FEDER-UE grants FPA2015-65929-P and
FPA2016-78645-P, by Maria de Maetzu program grant MDM-2014-0367 of
ICCUB, by the ``Severo Ochoa'' program grant SEV-2012-0249 of IFT, by
the Fermilab Graduate Student Research Program in Theoretical Physics
operated by Fermi Research Alliance, LLC, by the Villum Foundation
(Project No.~13164), and by the Danish National Research Foundation
(DNRF91).  Fermilab is operated by Fermi Research Alliance, LLC under
Contract No. DE-AC02-07CH11359 with the United States Department of
Energy.

\appendix

\section{Resolving LMA-D by COHERENT data}
\label{sec:app}

In this appendix we provide an analytic discussion of the ability to
resolve the LMA-D degeneracy using a combination of oscillation and
coherent-scattering data, \ie we focus on light NSI mediators.

In a coherent scattering experiment at a stopped pion source, two
combinations of $e$- and $\mu$-like events can be measured by using
timing information, see Sec.~\ref{sec:COHERENT}. Hence, an ideal
experiment would be able to extract both the electron- and
muon-neutrino scattering cross sections. Effectively the two parameter
combinations $Q_{we}^2$ and $Q_{w\mu}^2$ given in
Eq.~\eqref{eq:QwCOHERENT} can be measured. Let us set all off-diagonal
NSI to zero and assume that NSI happen either with up or down
quarks. Then we can write
\begin{align}
  Q_{w\alpha}^2 \propto (X_q - \eps^{q,V}_{\alpha\alpha})^2\,,
\end{align}
with
\begin{align}
  X_u = - \frac{Z g^{p,V} + N g^{n,V}}{2Z+N} \,,\qquad
  X_d = - \frac{Z g^{p,V} + N g^{n,V}}{Z+2N} \,.
\end{align}

We now introduce the following notation:
\begin{equation}
  \begin{aligned}
    s_q &= \eps_{ee}^{q,V} + \eps_{\mu\mu}^{q,V}\,,\\
    d_q &= \eps_{ee}^{q,V} - \eps_{\mu\mu}^{q,V}\,.
  \end{aligned}
\end{equation}
Oscillations are sensitive only to $d_q$, with best-fit $d_q\approx
0.3$, and the LMA-D degenerate solution at $d'_q = -d_q-\xi_q$
according to Eq.~\eqref{eq:NSI-deg}, with the $\xi_q$ given in
Eq.~\eqref{eq:xi}. COHERENT depends also on $s_q$. The transformation
$d_q\to d'_q$ can be supplemented by $s_q\to s'_q$. If we can arrange
$s_q$ and $s'_q$ such that
\begin{equation}
  \begin{aligned}
    \left[2X_q - (s_q+d_q)\right]^2 &= \left[2X_q - (s'_q+d'_q)\right]^2\,, \\
    \left[2X_q - (s_q-d_q)\right]^2 &= \left[2X_q - (s'_q-d'_q)\right]^2\,,
  \end{aligned}
\end{equation}
then COHERENT will be affected by the degeneracy as well. We use $d'_q
= -d_q-\xi_q$ and take the square root of the above equations. The two
non-trivial sign combinations lead to
\begin{equation}
  \label{eq:deg-conditions}
  \begin{aligned}
    2X_q - (s_q+d_q) &= \pm \left[2X_q - (s'_q-d_q-\xi_q)\right] \,, \\
    2X_q - (s_q-d_q) &= \mp \left[2X_q - (s'_q+d_q+\xi_q)\right] \,.
  \end{aligned}
\end{equation}

Using only information from oscillations, the sum $s_q$ is unknown. So
we can consider those equations for the two unknowns $s_q$ and
$s'_q$. In particular we find
\begin{equation}
  s_q = 2X_q \mp(d_q+\xi_q) \,,
\end{equation}
which implies
\begin{equation}
  \label{eq:deg-values}
  \eps_{ee}^{q,V} = \left\{
  \begin{aligned}
    & X_q - \xi_q/2 \\
    & X_q + \xi_q/2 + d_q
  \end{aligned}
  \right.
  \qquad\text{or}\qquad
  \begin{aligned}
    \eps_{ee}^{q,V} &= X_q - \xi_q/2 \\
    \eps_{\mu\mu}^{q,V} &= X_q + \xi_q/2
  \end{aligned} \,.
\end{equation}
Numerically we obtain
\begin{equation}
  \eps_{ee}^{u,V} =
  \left\{
  \begin{array}{r}
    -0.150 \\ 0.842
  \end{array}
  \right.
  \qquad
  \eps_{ee}^{d,V} =
  \left\{
  \begin{array}{r}
    -0.224 \\ 0.886
  \end{array}
  \right. \,,
  \label{eq:deg-values-num}
\end{equation}
where we used $d_u = 0.307, d_d = 0.316$ from the best-fit values for
oscillation data, and we took $Z=32, N=44$ for \Ge. The values for
$\eps_{ee}^{u,V}$ and $\eps_{\mu\mu}^{u,V}$ following from
Eq.~\eqref{eq:deg-values} are shown as the dashed lines in
Fig.~\ref{fig:2Dlight}.  If those are the true values for
$\eps_{ee}^{q,V}$ or $\eps_{\mu\mu}^{q,V}$, COHERENT will not be able
to resolve the degeneracy. Those estimates are in excellent agreement
with the numerical results shown in
Fig.~\ref{fig:LMA-D-light-future}. We can make the following comments:
\begin{itemize}
\item If there are no NSI in Nature ($d_q = s_q = 0$),
  Eqs.~\eqref{eq:deg-conditions} cannot be fulfilled, which implies
  that the degeneracy is resolved, in agreement with
  Fig.~\ref{fig:1Dlight}. This follows also from the fact that the
  dashed lines in Fig.~\ref{fig:2Dlight} do not pass close to the SM
  point.

\item If we use a reactor instead of the stopped pion source for the
  coherent scattering experiment, only the first equation
  in~\eqref{eq:deg-conditions} applies (corresponding to
  $Q_{we}^2$). For given $d_q$ and $s_q$ there is always a solution
  for $s'_q$. Hence, a reactor experiment combined with oscillation
  data cannot resolve the degeneracy in the light mediator case.

\item For a different target nucleus, the values of $X_q$
  change. Hence, Eqs.~\eqref{eq:deg-conditions} cannot be fulfilled
  simultaneously for two targets and in principle the degeneracy can
  be resolved by using different target nuclei.  However, as discussed
  in the context of Fig.~\ref{fig:LMA-D-light-future}, the effect is
  small and exploring it to resolve the degeneracy may be
  experimentally challenging.
\end{itemize}

\bibliographystyle{JHEP}

\providecommand{\href}[2]{#2}\begingroup\raggedright\endgroup

\end{document}